\pdfoutput=1
\documentclass[useAMS,usenatbib,usegraphicx]{mn2e}
\usepackage{amssymb,amsmath,amsfonts,graphicx}
\usepackage{epsfig,color} 

\usepackage{lipsum}

\newcommand{\apjl}{ApJL}
\newcommand{\apjs}{ApJS}

\newcommand{\aap}{A\&A}

\usepackage{times}
\usepackage[bookmarks,bookmarksnumbered,colorlinks=true, citecolor=blue, linkcolor=black]{hyperref}
\usepackage{color}

\newcommand{\blue}[1]{\textcolor{blue}{#1}}

\title[Cluster and Field Dwarfs in Illustris]{On the Assembly of Dwarf 
Galaxies in Clusters and their Efficient Formation of Globular Clusters}

\author[Mistani et al.]{
\parbox[t]{\textwidth}{
Pouria A. Mistani$^{1}$ $\thanks{E-mail: pakba002@ucr.edu}$,
Laura V. Sales$^{1,2}$,
Annalisa Pillepich$^{2}$,
Rub{\'e}n Sanchez-Janssen$^{3}$, 
Mark Vogelsberger$^{4}$, 
Dylan Nelson$^{2}$,
Vicente Rodriguez-Gomez$^{2}$,
Paul Torrey$^{4}$ and\\
Lars Hernquist$^{2}$
} 
\\
\\
  $^{1}$Department of Physics and Astronomy, University of California, Riverside, CA, 92521, USA\\
  $^{2}$Harvard-Smithsonian Center for Astrophysics, 60 Garden Street, Cambridge, MA, 02138, USA\\
  $^{3}$NRC Herzberg Institute of Astrophysics, 5071 West Saanich Road, Victoria, BC, V9E 2E7, Canada\\
  $^{4}$Department of Physics, Kavli Institute for Astrophysics and Space Research, Massachusetts Institute of Technology, Cambridge, MA 02139, USA\\ 
}
\begin{document}

\maketitle

\begin{abstract} 
  \noindent Galaxy clusters contain a large population of low mass
  dwarf elliptical galaxies whose exact origin is unclear: their
  colors, structural properties and kinematics differ substantially
  from those of dwarf irregulars in the field.  We use the Illustris
  cosmological simulation to study differences in the assembly
  histories of dwarf galaxies ($3\times 10^8 <\rm M_*/M_\odot<
  10^{10}$) according to their environment.  We find that cluster
  dwarfs achieve their maximum total and stellar mass on average $\sim
  8$ and $\sim 4.5$ Gyr ago (or redshifts $z=1.0$ and $z=0.4$,
    respectively),  around the time of infall
  into the clusters.  In contrast, field dwarfs not subjected to
  environmental stripping reach their maximum mass at $z=0$.
  These different assembly trajectories naturally produce a color
  bimodality, with blue isolated dwarfs and redder cluster dwarfs
  exhibiting negligible star-formation today.  The cessation of
  star-formation happens over median times $3.5$-$5$ Gyr depending on
  stellar mass, and shows a large scatter ($\sim 1$-$8 \; \rm Gyr$),
  with the lower values associated with starburst events that occur at
  infall through the virial radius or pericentric passages.  We argue
  that such starbursts together with the early assembly of cluster
  dwarfs can provide a natural explanation for the higher specific
  frequency of globular clusters (GCs) in cluster dwarfs, as found
  observationally. We present a simple model for the formation and
  stripping of GCs that supports this interpretation.  The origin of
  dwarf ellipticals in clusters is, therefore, consistent with an
  environmentally-driven evolution of field dwarf irregulars.
  However, the $z=0$ field analogs of cluster dwarf progenitors have
  today stellar masses a factor $\sim 3$ larger --a difference arising
  from the early truncation of star formation in cluster dwarfs.
\end{abstract}


\section{Introduction}
\label{sec:intro}

By number, dwarf elliptical galaxies (dEs) dominate the population of
dwarfs in cluster environments. They are, however, almost absent in
the field, where the prevalent morphological type is gas-rich dwarf
irregulars (dIrrs) \citep{Binggeli1990, Grebel1999, Geha2012}.  This
difference immediately raises questions about the relationship between
dEs and dIrrs.  Are these two types of galaxies intrinsically
different objects?  Or, is there an evolutionary link between these
two dwarf populations?

There is no strict definition for dE galaxies, but they are rather
characterized by a combination of attributes such as having low mass,
being of low surface brightness, and exhibiting red colors and low
levels of star formation. Moreover, dEs have quite a heterogeneous
morphological mix \citep{Lisker2009, Janz2012} and also show wide
kinematic variations (\blue{Geha et al. 2002; van Zee et al. 2004a;
Lisker2006;  Ry{\'s} et al. 2013; Toloba et al. 2014a,b}).
These observations hint at a complex formation scenario where more
than one factor determines their final fate. Although the definition
is not specific, there is clear consensus that dEs are less-disky and
less rotationally-supported than dwarf irregulars in the field.
\nocite{Geha2002, vanZee2004, Lisker2006, Rys2013,Toloba2014a,Toloba2014b}

Differences between the two populations could arise from either
internal or external mechanisms.  It appears that dIrrs and dEs follow
similar scaling relations, so they could be the same type of object
but in which internal processes, like star-formation and feedback,
caused some to retain more gas than others \citep{Dekel1986,
  deRijcke2005}.  While appealing, such a scenario cannot explain the
strong correlation observed with the environment. Alternatively, dEs
may have been dIrrs in the past that were transformed into dEs later,
due to environmental effects within their host clusters. Such a
transformation could involve some degree of ram-pressure stripping and
gas starvation to redden their colors (\blue{Conselice et al. 2003, van 
Zee et al. 2004b, Boselli et al. 2008}), and tidal effects to heat them
kinematically, altering their morphologies \citep{Mayer2001,
  Gnedin2003, Lisker2009, Donghia2009, Smith2010,Lelli2014}.  A different and
intriguing perspective for the influence of environment on dwarfs was
argued by \citet{Sabatini2005}, who noted that interactions with the
cluster medium could temporarily {\it enhance} star formation in
infalling dIrrs, leading to rapid consumption of the gas and a
transformation into dEs.  Finally, low mass dEs could also be the
stripped remnants of more massive galaxies that fell into the cluster
long ago \citep{Conselice2003a}.
\nocite{Conselice2003a,vanZee2004b, Boselli2008}

Environmentally driven transformation models for dEs, although
attractive, are not free from challenges. One of the main problems
with this scenario is the observed difference in the frequency of
globular clusters (GCs) between dwarfs in clusters and in the
field. If dEs and dIrrs are evolutionarily linked, dEs should contain
similar or even fewer GCs than disky galaxies in the field because of
tidal stripping within the cluster host.  Instead, dEs in
clusters tend to show an {\it increased} GC specific frequency over
dIrrs at similar masses (\blue{Miller et al. 1998, Miller \& Lotz 2007, 
Jordan et al 2007b, Peng et al. 2008, Sanchez-Janssen \& Aguerri 2012}). 
This discrepancy appears accentuated
for nucleated dEs \citep{MillerLotz2007} and dwarfs in the inner
regions of clusters (\blue{Jordan et al. 2007b, Peng et al. 2008}).
\nocite{Miller1998, MillerLotz2007, Jordan2007b,Peng2008, Sanchez-Janssen2012}

With their capability of being able to track the evolution of
galaxies, numerical simulations could, in principle, resolve these
issues.  In practice, this is difficult because GC formation is
still poorly understood cosmologically (however see an original
attempt by \citealt{Prieto2008}), and the dynamic range needed to
self-consistently evolve dwarf galaxies with their GC systems
in a galaxy cluster environment is currently prohibitive.
Here, we propose a hybrid approach.  Hydrodynamical simulations of a
large cosmological volume are performed to enable us to sample a set
of galaxy clusters and, simultaneously, to follow the assembly and
evolution of dwarfs in clusters and in the field.  In post-processing,
we then apply a simplistic model of GC formation together with a
particle-tagging technique to roughly describe the formation and
stripping of GCs in cluster dwarfs.  Our effort builds on the earlier
work by \citet{Peng2008} based on semi-analytical catalogues and
incorporates the effect of tidal stripping of GCs not addressed in
their work.

This paper is organized as follows. We introduce the simulations and
selection criteria for our samples in Sec.~\ref{sec:simul}. The
assembly of dwarfs according to their environment is investigated in
Sec.~\ref{sec:assembly} and the details of their star formation
histories is presented in Sec.~\ref{sec:sfr}. We develop and apply our
method for tracking GCs in Sec.~\ref{sec:GC}, addressing the expected
GC frequencies for cluster and field dwarfs. We summarize our main
results in Sec.~\ref{sec:concl}.

\begin{center} \begin{figure*} 
\includegraphics[width=0.9\linewidth]{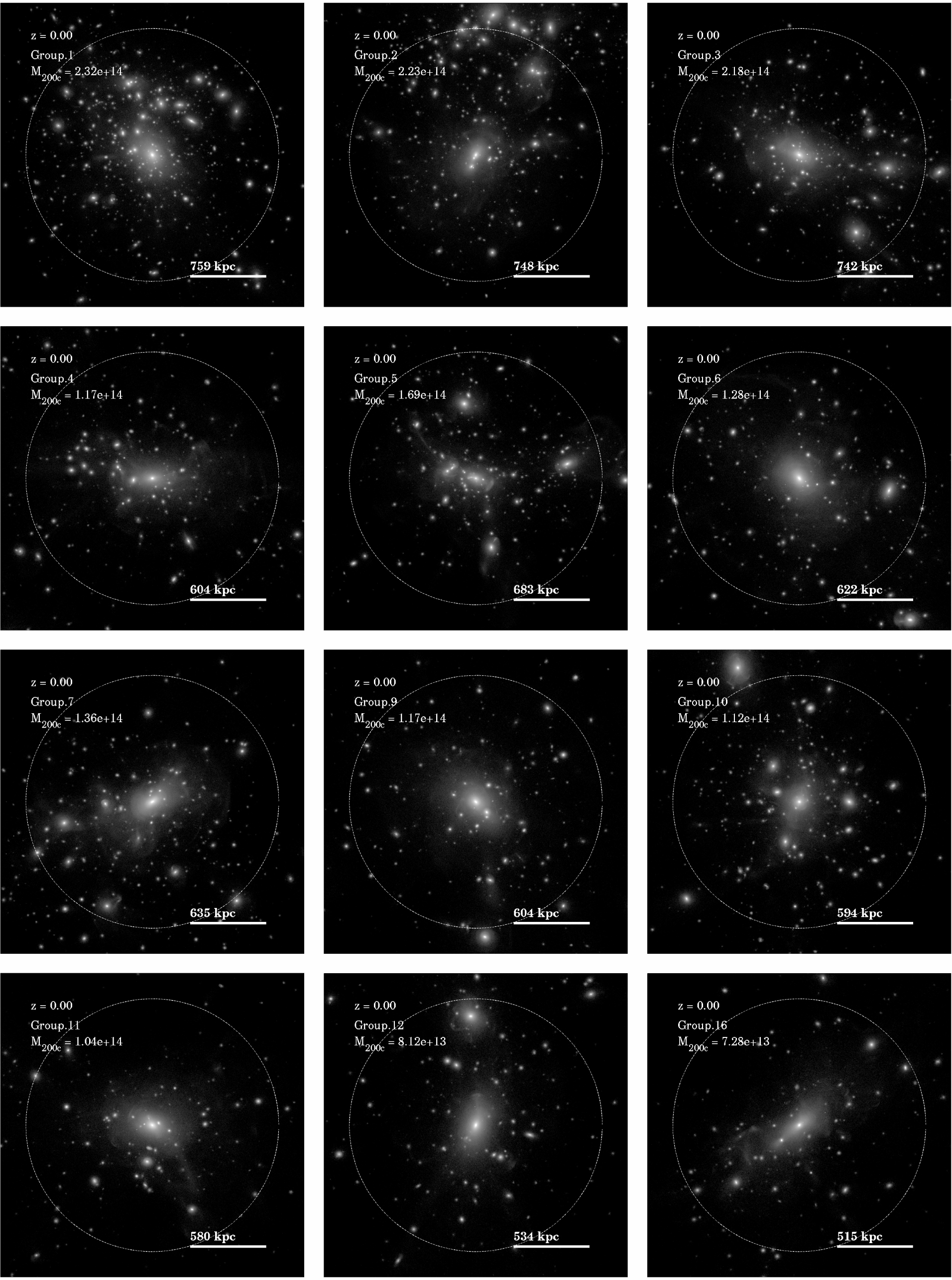}
\label{clusters}
\caption{Projected stellar maps for the $12$ most massive galaxy
  clusters in Illustris ($M_{200}\ge 5 \times 10^{13}\; \rm M_\odot$). Individual ``clumps''
  correspond to satellite galaxies inhabiting each of the
  clusters. Circles indicate the extent of the virial radius.}
\label{fig:image}
\end{figure*}
\end{center}

\section{Numerical Simulations}
\label{sec:simul}

\begin{center} \begin{figure*} 
\includegraphics[width=0.49\linewidth]{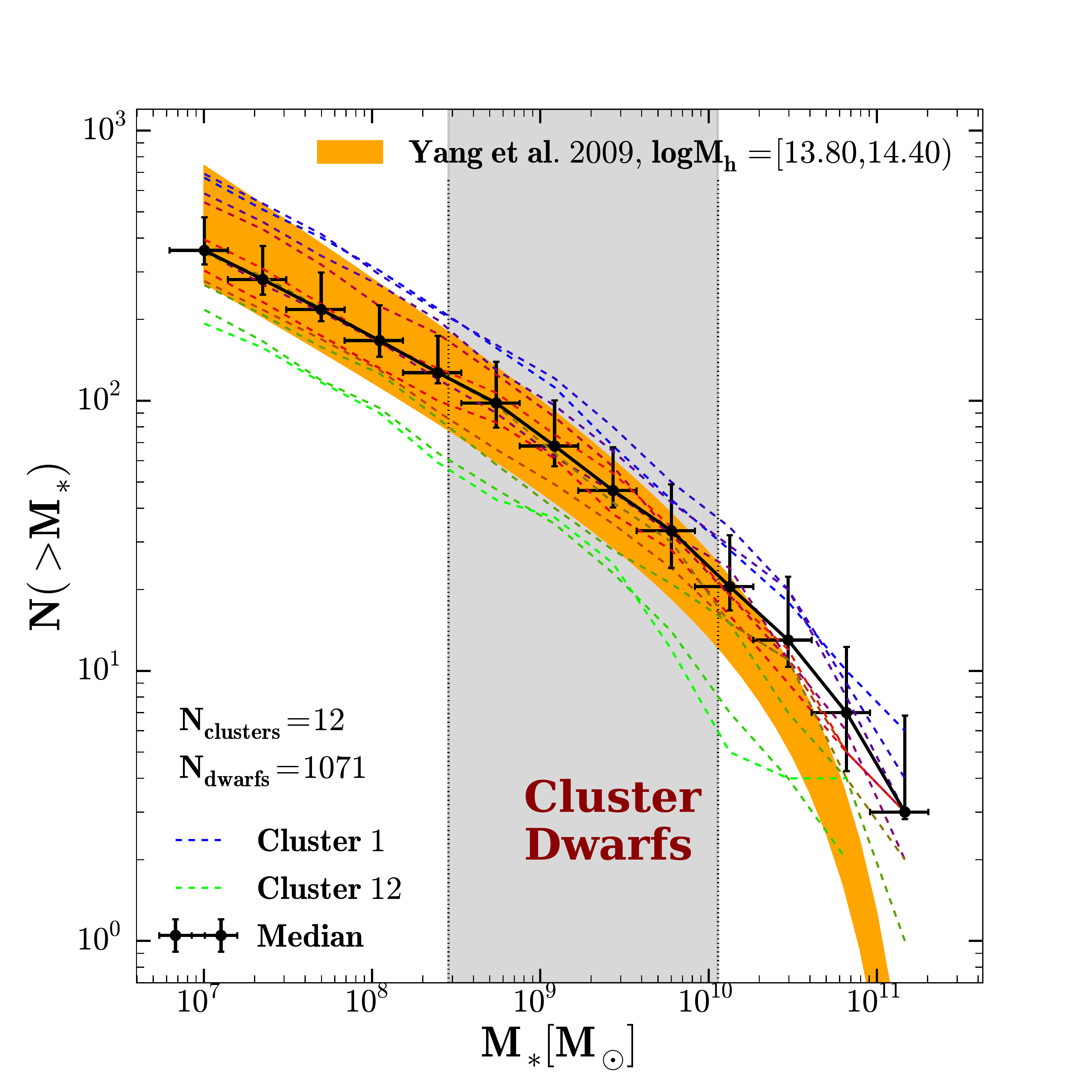}
\includegraphics[width=0.49\linewidth]{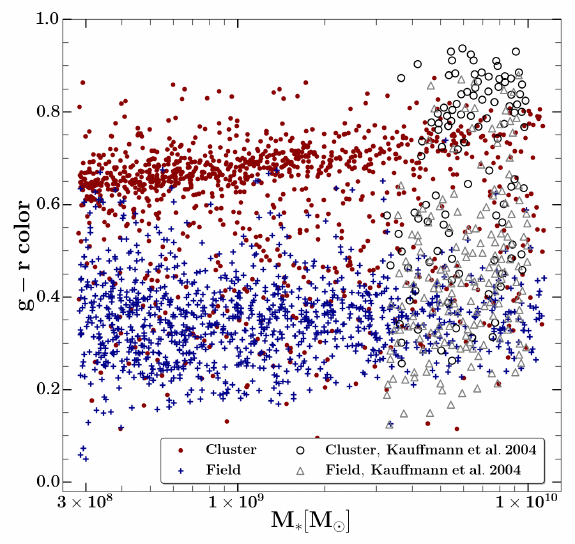}
\caption{Left: Cumulative stellar-mass function of satellite galaxies
  for the simulated clusters in our sample. Individual clusters are
  shown with colored dashed lines and the median is indicated with
  solid black line/symbols. Our selection criteria for dwarf galaxies
  ($M_*=[3 \times 10^{8}-1 \times 10^{10}]\; \rm M_\odot$) is
  highlighted by the gray area. We find a total of 1071 cluster
  dwarfs. Note the good agreement with the SDSS observed satellite
  mass function (taken from \citet{Yang2009}, orange shading),
  particularly in the selection region of our dwarfs. Right: The
  color-stellar mass distribution of dwarfs in our cluster sample
  (red dots) and field control sample (blue crosses), selected to be
  central (not satellites) objects in the same mass range.  Encouragingly,
  cluster dwarfs are much redder than the field sample, 
  as indicated by observations.  We show in grey circles/triangles
  SDSS results for galaxies in low/high density environments taken
  from \citet{Kauffmann2004}.}
\label{fig:sample}
\end{figure*}
\end{center}

We employ a set of model galaxies taken from the Illustris
Simulation (\blue{Vogelsberger et al. 2014a,b}; \blue{Genel et
  al. 2014}). Illustris consists of a cosmological box $106.5$
Mpc on a side evolved with hydrodynamics using the code {\sc arepo}
\citep{Springel2010}. The simulation adopts a
WMAP9-consistent cosmology and includes treatments of gravity together with
the most relevant physical processes driving galaxy evolution,
including 
gas cooling/heating \citep{Katz1996, Faucher2009},
star-formation \citep{SH2003}, metal enrichment, and AGN
feedback \citep{SDH2005, DSH2005, Sijacki2007}.
While a detailed description of the model can be found
elsewhere \citep{Vogelsberger2013}, for completeness we
briefly summarize the relevant aspects of the physics included in our
simulations. \nocite{Vogelsberger2014a, Vogelsberger2014b}

Gravity is computed using a Tree-PM approach in which long-range
forces are calculated on a grid while short-range interactions are
obtained by expanding the potential into multipoles
\citep{Springel2005}. The Euler equations governing the hydrodynamics
of the gas are solved on an unstructured mesh that moves with the
flow, yielding a continuously adaptive method to achieve high
resolution where needed and simultaneously moderating the impact of
numerical diffusion \citep{Vogelsberger2012, Sijacki2012, 
Genel2013, Nelson2013}.
In this scheme, the gas is distributed among ``cells '', which can be
heated or cooled according to the local density and metallicity.
Dense gas becomes eligible for star-formation above a density
threshold $n_H=0.13 \; \rm cm^{-3}$, at which point star particles are
spawned probabilistically according to the local dynamical time.

We follow the evolution of stars, including mass and metal return from
stellar winds and the explosion of massive stars (supernova type I and
II; hereafter, SNI and SNII, respectively). In the case of SNII, we
also consider the deposition of kinetic energy into galactic-scale
winds, with initial velocities that vary linearly with the local
velocity dispersion of the dark matter halo. The mass loading is then
determined from the available energy from SNII explosions \citep{SH2003}.
With these choices for the sub-resolution physics, the
Illustris simulations have been shown to reproduce several of the
observational properties of galaxies, such as their stellar mass and color
distribution, morphological variety, angular momentum distribution
and typical scaling relations, among others \citep{Torrey2014,
Genel2014, Genel2015, Snyder2015, Sales2015}.

In the highest resolution run, or Illustris-1 (used here), the mass
per DM and baryonic particle are $m_{\rm DM} \sim 6.26\times 10^{6} \;
M_\odot$ and $m_{\rm b} \sim 1.26\times 10^{6} \; M_\odot$,
respectively.  Gravity is softened with a spline kernel \citep{HK1989}
and the softening length
scales with redshift but is always comparable to or better than
$\epsilon=700(1400)$ pc for the stellar (dark matter) particles. The
most well-resolved gas elements can have extents as small as  $48$ pc.

Galaxies are identified as self-bound structures at each output using
a combination of a Friends-of-Friends (FoF) algorithm
\citep{Davis1985} for groups and a subsequent application of the {\sc
  subfind} algorithm \citep{Springel2001, Dolag2009}.  The centers are
marked by the position of the particle with the minimum potential
energy and galaxy properties are computed within twice the half-mass radius
of the stars for each object. The time evolution of a given object is
tracked by means of {\sc LHaloTree} merger trees \citep{Springel2005N,
  Nelson2015} from a total of 136 output snapshots covering the
redshift range $z=50$ to $z=0$.

\begin{center} \begin{figure*}
    \includegraphics[width=190mm]{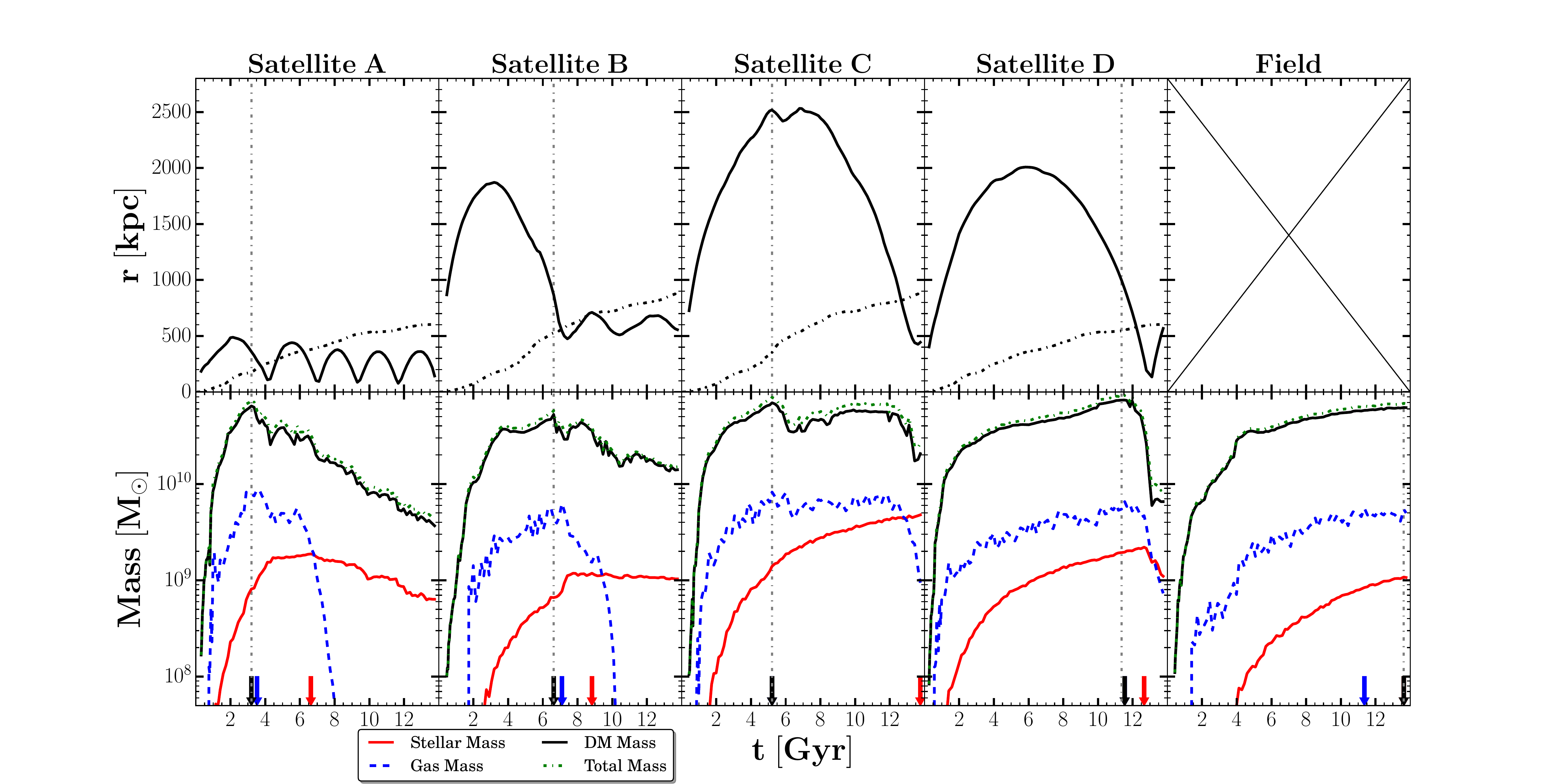} 
    \caption{Examples of the orbits (top row) and mass evolution
      (bottom row) of cluster dwarfs (Satellites A-D) and a field
      dwarf (rightmost column). Orbits show a wide diversity, with
      some dwarfs completing more than $3$-$4$ revolutions around the
      cluster (example Sat. A) and others only recently arriving
      (Sat. D). The time evolution of the host cluster's virial radius $r_{200}$
        are indicated with a black dashed line in each panel. 
      The time of maximum {\it total} mass (see green-dotted line
      in the bottom row) is indicated in all panels with a vertical dotted line. 
      This time corresponds roughly to the moment they 
      stopped being centrals to become satellites; which can happen 
      right before crossing the virial radius of the cluster 
      like in Sat. A, B and D, or 
      before, if they were accreted into a smaller group first and then
      entered the cluster (like Sat. C). The mass evolution in the bottom
      row shows clear correlations with the orbits, with a decrease of mass after
      infall as well as close pericentric passages. Tidal stripping
      is not strong for the stellar component (solid red), although
      dwarfs in tightly bound orbits can experience significant stellar mass
      loss (example Sat. A). Small vertical arrows show the times at
      which each component (dark-matter, stars, gas) reaches
      its maximum. Stars continue to build up after infall, as shown
      by the shift between the dotted vertical line and the red
      arrows. Note that for field dwarfs, which are not exposed to stripping,
      the mass in all components peaks only at the present time.}
\label{fig:orbits} \end{figure*}
\end{center}

\subsection{Simulated Dwarfs in Clusters and in the Field}
\label{sec:sample}

Our sample of dwarfs consists of a set of ``cluster'' dwarfs selected
to inhabit the most massive groups and clusters in {\it Illustris},
and ``field'' dwarfs, a control sample consisting of central low mass
galaxies that are isolated.  The selection is done once at $z=0$
  and the time evolution of both populations is studied by following
  the progenitors of each object through time. To construct the
cluster sample, we consider all galaxy clusters with virial mass
$M_{200} \ge 5 \times 10^{13}$ $\rm M_\odot$, where the virial radius
$r_{200}$ is defined to be that which encloses $200$ times the
critical density of the Universe, and the virial mass $M_{200}$ is the
total mass contained within $r_{200}$.  Applying these criteria, we
obtain a total of $12$ clusters, with the most massive reaching
$M_{200}=2.32\times 10^{14} \; \rm M_\odot$, similar to mass estimates
for the Fornax (\blue{Jordan et al. 2007a}) and the Virgo
\citep{Bohringer1994} galaxy clusters.

In Fig.~\ref{fig:image} we show stellar density maps of the clusters
in our sample, with the virial mass and original FoF-ID quoted in each
panel. All clusters are populated by a large number of {\it satellite}
galaxies, seen as individual ``clumps'' in each image, which cover
a wide range of masses and properties. By definition, satellite
galaxies are all objects identified by {\sc subfind} that are not the
centrals of their FoF group.

The left panel of Fig. ~\ref{fig:sample} shows the cumulative
stellar-mass function of these satellite galaxies, indicated by
colored dashed lines for each of the selected $12$ clusters. We
  consider only satellites that are at $z=0$ within the virial radius
  of their host system. The median of our sample (black symbols/solid
line) agrees well with the satellite stellar-mass function of groups
and clusters of similar mass in SDSS, as indicated by the orange
shaded region \citep{Yang2009}. Cluster dwarf galaxies are then
selected from these satellites by a simple mass cut, $3\times 10^8 <
M_*/M_\odot < 1\times 10^{10}$ (gray region), implying that all our
dwarfs are resolved with at least $240$ stellar particles today. This
selection criterion yields a total of 1071 cluster dwarfs.

For comparison, we also define a sample of ``field'' dwarfs in the
same stellar mass range but that are {\it central} galaxies of their
FoF group (i.e. they are not satellites of any larger system).  A
total of 28391 dwarfs satisfy this condition, from which we randomly
select 1071 to match the number of cluster dwarfs.  Encouragingly,
this simple selection procedure results in interesting differences
between the color distributions of the ``cluster'' and ``field'' dwarf
populations, as expected from observations of galaxies in different
environments. The right panel of Fig.~\ref{fig:sample} shows that at
fixed stellar mass, the cluster dwarfs (red dots) are significantly
redder than the field objects (blue crosses), in good agreement with
the observed redder colors of satellites around massive hosts
\citep[e.g., ][]{Dressler1980,Martinez2002,Wang2014}.  This suggests
that our simulation properly captures the main physical processes
driving the evolution of galaxies in different environments and can
therefore shed light on the formation and evolution of dEs in cluster
environments.

\section{The assembly of dwarf galaxies in different environments }
\label{sec:assembly}
%
 \begin{center} \begin{figure}
     \includegraphics[width=84mm]{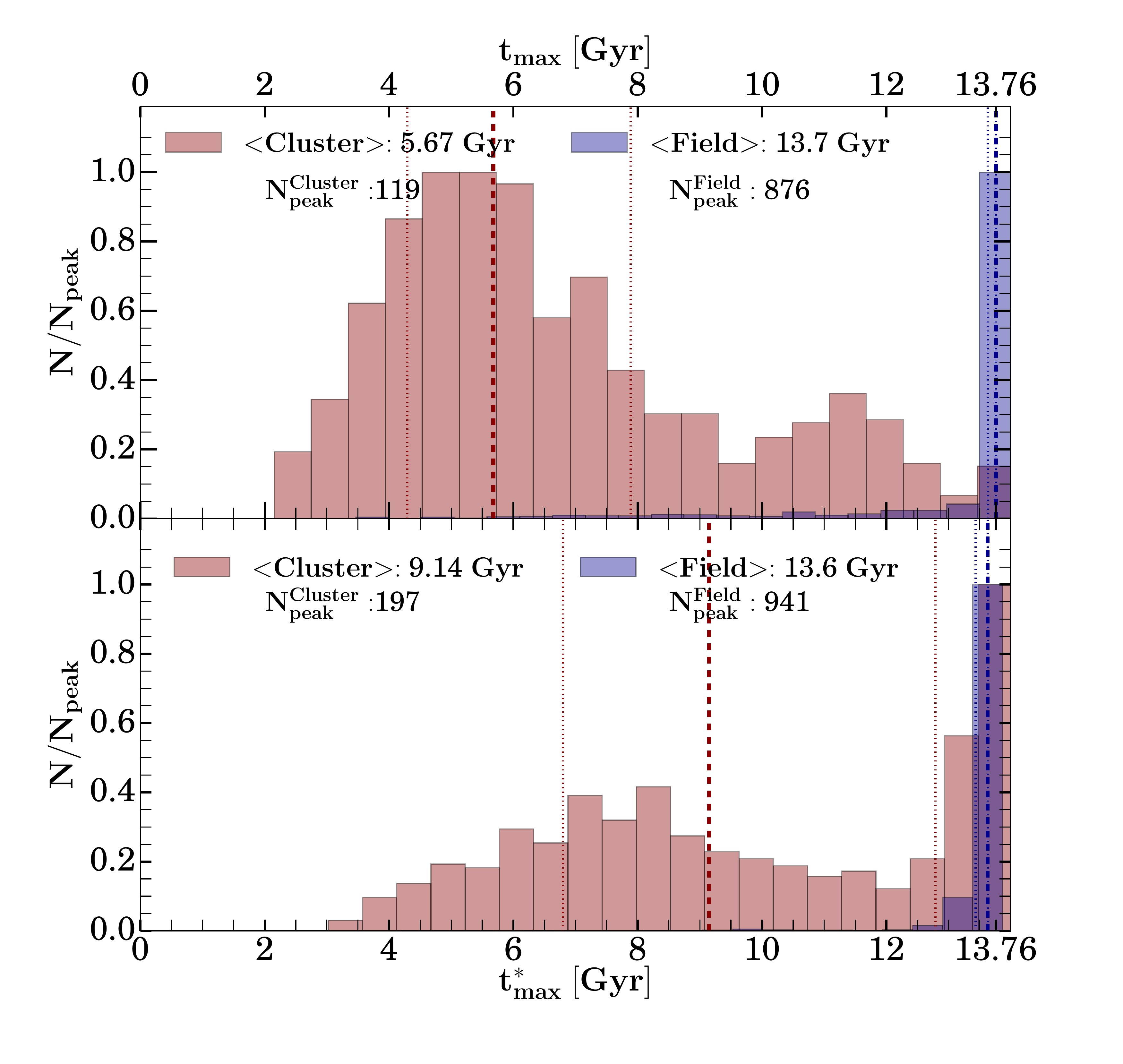} 
     \caption{Distributions of the time of maximum {\it total} mass
      $t_{\rm max}$ (top) and maximum {\it stellar} mass
       $t_{\rm max}^*$ (bottom).  Histograms are normalized to the peak 
       value in each sample (values quoted). Cluster/Field dwarfs are shown in
       red/blue. Due to stripping, dwarfs in clusters assemble their
       mass earlier, typically reaching their maximum mass at
       $t_{\rm max} \sim 5.5$ Gyr or $z \sim 1$, unlike the field population
       which continues to grow until today ($t_{\rm max} \sim 13.7$
       Gyr). The time of maximum {\it stellar} mass in
       cluster dwarfs is about $3.5$ Gyr later than the time of
       maximum {\it total} mass, indicating that satellite dwarfs can
       continue to form stars for a significant period of time after
       they have become satellites.}
\label{fig:tmax}
\end{figure}
\end{center}

The cluster dwarfs sample a wide variety of orbits around their
central hosts. We show several examples in the top row and first 4
columns of Fig.~\ref{fig:orbits}. Dwarf satellites initially follow
the Hubble expansion until they are captured by the gravitational
field of a cluster and then turn around (seen as the moment of maximum
distance from the cluster). They then begin the process of infall and
settle into their orbits around the host potential. Some dwarfs
in our sample have fallen in more than $10$ Gyr ago, completing
several revolutions around the cluster's center by the present
(e.g. Satellite A). On the other hand, some objects have only
recently been accreted into the cluster (Satellite D), crossing the host
virial radius less than $2$ Gyr ago (the time evolution of the host
virial radius is shown with a dashed line).  From this perspective,
the sample of cluster dwarfs is a complex group of objects whose
properties are expected to be diverse in light of the variations
present in their orbital histories.

Orbits have a clear impact on the mass evolution of cluster dwarfs, as
shown in the bottom row of Fig.~\ref{fig:orbits}.  Dwarfs initially
grow as expected (green dotted line shows the evolution of total mass)
but they later reach a maximum after which their total mass decreases
due to tidal stripping. The maximum is usually reached at the last
time the dwarfs are the central object of their FoF group. Therefore,
this time of maximum mass, shown as vertical dotted lines in each
panel, is a good indicator of the time of infall, or the time when
dwarfs stopped being central objects to become satellites.  In what
follows, we will refer to the time of infall $t_{\rm inf}$ or infall
redshift $z_{\rm inf}$ as the time when the simulated dwarfs reached
their maximum total mass $t_{\rm max}$.

There are two interesting points to notice from these growth curves.
Some dwarfs (e.g. Satellite C) first become a satellite of a different
system that later fall into the cluster. This pre-processing of
dwarfs is not uncommon, and is usually accompanied by a wiggly pattern
in their orbits as they move around a different host on the way to
their final infall into the cluster (see Satellite C around $t\sim
5.5$ $\rm Gyr$).  Pre-processing effects in groups have been
hypothesized to account for the observed properties of the Magellanic
Clouds \citep{Besla2007, Besla2010, Besla2012}, and is also expected
to happen regularly in cluster dwarfs \citep[e.g. ][]{Lisker2013}.
For our sample, $65\%$ of the objects infall first into an
intermediary system, where they spend a rather short amount of time
(median of $\sim 1.3$ Gyr), before finally infalling into their host
galaxy clusters where they stay until $z=0$.  Second, the small
difference between the green dotted and black solid lines suggests
that stripping mainly affects the dark matter component (solid black
line). The times when the maximum total mass $t_{\rm max}$ and maximum
dark matter mass $t_{\rm max}^{DM}$ are reached always track one
another (see black arrows).

Baryons, instead, show a different behaviour. Stars are much more centrally
concentrated than the dark matter component, resulting in a small
fraction of the stellar mass being tidally stripped (with some
exceptions as in Satellite A). In our sample of 1071 cluster dwarfs,
$62\%$ retain more than $80\%$ of their maximum stellar mass and
only $1\%$ have lost more than $90\%$ of their stars by $z=0$. For
comparison, this corresponds to $2\%$ and $18\%$ objects respectively
for the same thresholds in the dark matter component. Moreover, the
stellar mass continues to grow after dark matter stripping has
begun, as indicated by the red arrows marking the time of maximum
stellar mass for each object, $t_{\rm max}^*$.  This agrees well with
previous findings in Illustris of significant star formation continuing after
satellite infall \citep{Sales2015, Vicente2015}. The
gaseous component (dashed blue line) stops growing around the time of
infall and tends to be equally or more greatly affected by the
environment than the dark matter component. Tidal stripping as well as
gas consumption from star formation and hydrodynamical
effects like ram-pressure stripping contribute to this effect. Most cluster
dwarfs ($74\%$) show significant gas depletion, losing more than
$90\%$ of their maximum gas mass, which is consistent with the low
levels of star-formation driving the red colors seen previously in
Fig.~\ref{fig:sample}.

The evolution of dwarfs in the field is remarkably different. The last
column in Fig.\ref{fig:orbits} shows the mass evolution of a random
dwarf from the field sample.  Of note, all components show a steady
mass growth until $z=0$, with occasional pronounced mass-gain
events associated with mergers (for example at $t\sim 4$ Gyr). Galaxies
in the field do not experience significant stripping, and therefore,
reach a maximum mass at the present day, as shown by the vertical dotted
line and small arrows. This is not a feature of this particular dwarf,
but rather the norm in the whole field dwarf sample, as shown in
Fig.\ref{fig:tmax}. The top panel shows the distribution of $t_{\rm
  max}$ (the times when the total maximum mass is reached) for all
cluster (red) and field (blue) dwarfs. The median and $25\%$-$75\%$
percentiles are shown as vertical dashed and dotted lines,
respectively. For most dwarfs in the field, $t_{\rm max}\sim 13.7$ Gyr
(i.e., `today') whereas the median for cluster dwarfs is $t_{\rm max} \sim 5.6$
$\rm Gyr$ or $z=1.1$. A similar behaviour is seen for the time of
maximum stellar mass, with field dwarfs assembling their stars
significantly later than their cluster counterparts ($t_{\rm max}^{*}\sim
13.6$ and $9.14$ $\rm Gyr$ for field and cluster dwarfs
respectively). Fig.~\ref{fig:tmax} highlights that a significant
level of star formation after infall is a common feature for cluster
dwarfs (the median $t_{\rm max}$ and $t_{\rm max}^*$ differ for more than $\sim
3$ Gyr); a subject we will return to in Sec.~\ref{sec:sfr}.

\begin{center} \begin{figure} 
\includegraphics[width=85mm]{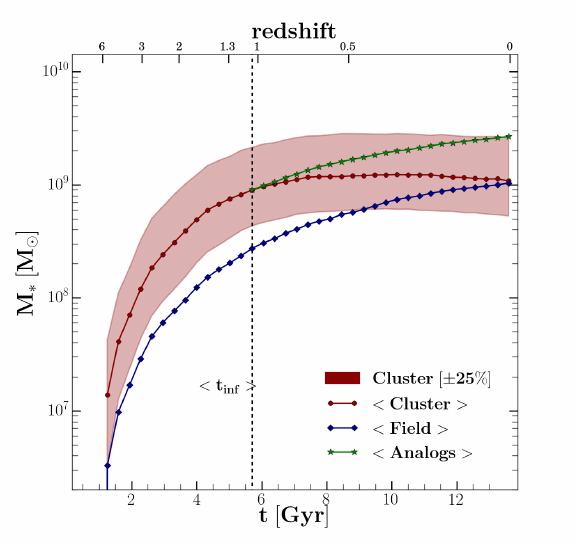}
\caption{Median stellar mass evolution for the sample of cluster (red)
  and field (blue) dwarfs. The $25\%$-$75\%$ dispersion on the median
  is shown as the shaded area for the cluster dwarf sample only, but
  is of similar magnitude for the field dwarfs. Although both samples
  have (by construction) the same mass at $z=0$ their time evolution
  is significantly different: cluster dwarfs were more massive than
  the field sample at earlier redshifts. Both samples gain mass  at
  approximately the same rate up to $z \sim 1$, when the growth of
  cluster dwarfs decelerates due to environmental effects (median
  infall time of the sample shown by the vertical dashed line). 
  In contrast, a sample of dwarf analogs that had the same
  properties as the progenitors of cluster dwarfs at 
{\it the time of infall} (instead of at z=0) would outgrow the median stellar
  mass in cluster dwarfs by up to a factor $\sim 3$ today (green
  line).  }
\label{fig:mstr_t}
\end{figure}
\end{center}

This different stellar mass assembly can be better appreciated from
Fig.~\ref{fig:mstr_t} for the population of cluster (red) and field
(blue) dwarfs.  Curves correspond to the median of the samples, and
the $25\%$-$75\%$ dispersion is indicated by the shaded region (for
clarity only shown for the cluster dwarfs but is of similar magnitude
for the field objects). Because of our selection, the curves overlap
at $z=0$ where both samples were chosen to be within the same stellar
mass range: $3\times 10^8 < M_*/M_\odot < 1\times 10^{10}$. However,
going back in time, the samples diverge. Field dwarfs form
their stars steadily throughout their history up to $z=0$. In contrast,
dwarfs in clusters grow their mass rapidly at early times, then being
$3$-$4$ times more massive than the field sample. However, by the
median time of infall of the cluster sample, $\left< t_{\rm inf} \right> \sim 5.5$
$\rm Gyr$ (vertical dashed line), the stellar mass freezes and does not appreciably
change thereafter.  This behaviour is caused by more dwarfs becoming
satellites, lowering their star-formation rates and being exposed to
tidal stripping.

This excess of mass at early times with respect to today's field
objects suggests that, if they were in the field, cluster
dwarfs would have grown significantly more in mass by
today. Processes associated with the cluster environment have
truncated that growth.  We examine this possibility by defining a group of
cluster dwarf ``analogs'' (green line), selected to have similar mass as the
cluster dwarfs {\it at time of infall} but that, unlike the
cluster dwarfs, remain isolated field objects until
the present\footnote{In practice, the selection of ``analogs'' is done as 
follows: at the median infall time of the cluster dwarfs sample 
$\left< z \right> \sim 1.1$, we select all field objects that 
at that time have a stellar mass within a factor of $2$ of the 
median stellar mass of the progenitors cluster dwarfs at the same time. 
We then follow the selected objects forward in time to
identify those that remain central galaxies to $z=0$ and choose
1071 objects randomly from that set.}. This exercise
shows that a sample selected to match cluster dwarfs at {\it infall}
rather than at $z=0$ ends up a factor $\sim 2.5-3$ more massive on
average than the cluster dwarfs today.  The most appropriate 
comparison sample is thus unclear 
and may depend on the question being addressed.
Our results indicate that selecting isolated objects in
the field having $\sim 3$ times more mass than the cluster dwarfs
we are targeting may provide a better match to the early phases of growth
of both samples.  To facilitate the comparison with observations, in
what follows we will still consider our ``field'' objects
selected within the same mass range as cluster dwarfs at $z=0$
to be the control sample, but we
bear in mind the lesson learned from the ``analogs'' sample.

%
\begin{center} \begin{figure} 
\includegraphics[width=85mm]{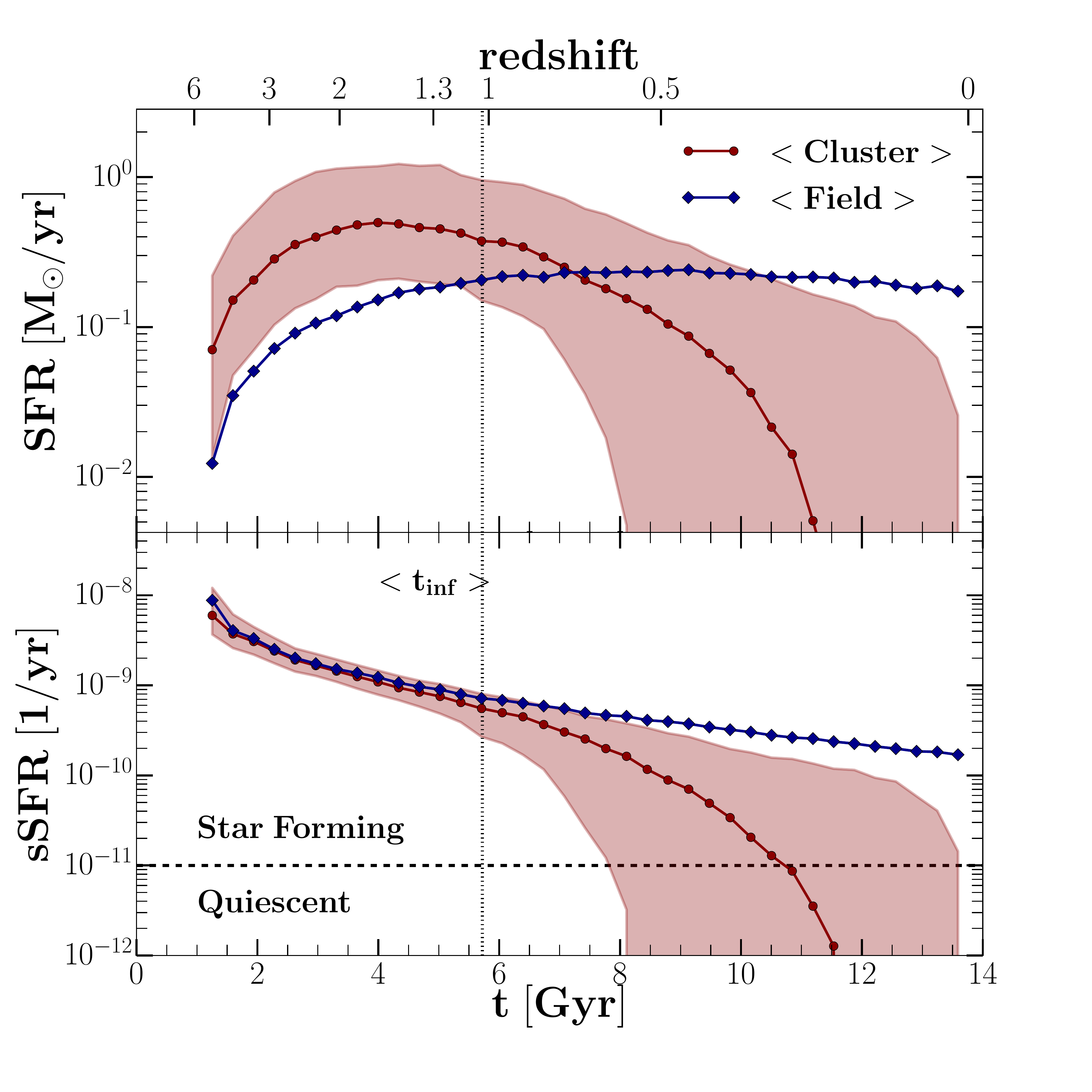}
\caption{Median star formation rate (SFR, top panel) and the specific star
  formation rate (sSFR, bottom panel) as a function of time for cluster (red) and field
  (blue) dwarfs. Color coding is the same as in Fig.~\ref{fig:mstr_t}. 
  Dwarfs in clusters form their stars at higher SFR than the field
  sample. However, the difference is only due to their larger stellar
  mass at high redshifts, since the sSFR of both samples agrees well
  before the average time of infall $\left<t_{\rm inf}\right>\sim 5.5$ Gyr. After
  that, the median rate of star formation in cluster dwarfs decreases, with
  most objects becoming quiescent by $z=0$. We define $sSFR=10^{-11}\;
  \rm yr^{-1}$ as our threshold to divide star-forming from quiescent
  objects (black dashed line). }
\label{fig:sfr_t}
\end{figure}
\end{center}
%
\section{The star formation history of dwarf galaxies: cluster vs. field }
\label{sec:sfr}

We now turn our attention to {\it how} dwarfs form their stars
according to their environment. Fig.~\ref{fig:sfr_t} shows the median
star formation rate (SFR, upper panel) and specific star formation
rate (sSFR, lower panel) of our samples as a function of time.
  We define the sSFR as the SFR divided by instantaneous stellar mass
  of the galaxy. As before, red/blue corresponds to cluster/field
dwarfs and $25\%$-$75\%$ percentiles are indicated by the shadowed
region. As expected from the mass difference seen in
Fig.~\ref{fig:mstr_t}, cluster dwarfs form stars at a higher rate than
in the field at early times, but the trend reverses as more dwarfs
become satellites after $z=1$.  By the present day, the vast majority
of cluster dwarfs have stopped forming stars, consistent with the red
colors displayed in Fig.~\ref{fig:sample}.  Interestingly, the SFR
difference at high redshifts is only due to the different masses of
the samples: once the SFR of each dwarf is normalized by their
(instantaneous) $M_*$, the specific rates between the samples are
consistent with each other (bottom panel). Cluster dwarfs are not
``special'' objects in this regard and simply form stars according to
the expected rate given their stellar mass.

\begin{center} \begin{figure*}
    \includegraphics[width=190mm]{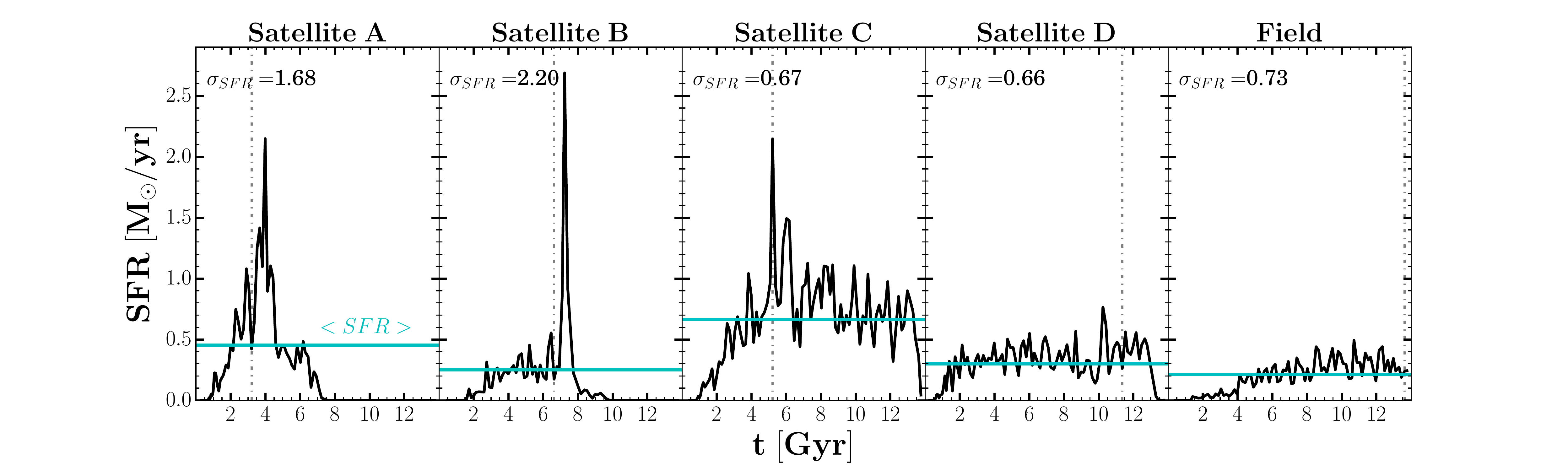} 
    \caption{Examples of individual star formation histories for the same
      satellite and field dwarfs in Fig.~\ref{fig:orbits}. As in the
      case of the orbits, the SFR histories of cluster dwarfs exhibit large
      variations. It is not uncommon to see ``peaks'' or episodes of
      intense star formation (for example Sat. A, B) that coincide
      with the crossing of the virial radius of the cluster and/or
      first pericentric passages. Starburst events are not common
      for field dwarfs. Some cluster dwarfs show a more constant SFR
      that is comparable to field objects (for example Sat. D). In
      each panel, we quote the ``starburstiness'' of the curve, 
      $\sigma_{\rm SFR}$, defined as the standard deviation of the
      SFR history with respect to the time-average $\left<\sigma_{\rm SFR}\right>$
    (cyan line). Large values of  $\left<\sigma_{\rm SFR}\right>$ indicate the
    presence of starbursts. (See text for more details.)} 
 \label{fig:sfr_panels} 
\end{figure*}
\end{center}

Beyond averaging the samples, examination of individual star
formation histories of dwarfs reveals another interesting aspect of
cluster dwarfs. Fig.~\ref{fig:sfr_panels} shows the SFR evolution
corresponding to Satellites A-D and the field dwarfs shown in
Fig.~\ref{fig:orbits}. The curves in the first three panels show
significant peaks or ``bursts'' of star formation that are not present
in the field objects or in Satellite D. Such bursts, when present, seem to occur
around the time of infall (indicated by the vertical dotted line) and
we have checked that these events are usually associated with
either the crossing of the virial radius of the cluster or to the time
of first pericentric passage. Often times the star formation is
quenched immediately following this final ``burst''. 

We would expect that the tidal and ram-pressure interactions with a cluster
could trigger a compression of the interstellar medium of a cluster
dwarf, temporarily raising its star formation rate. Such an effect has
previously been seen in idealized simulations
\citep{Barnes1996, Mihos1996, Mayer2001}.  Here we
find that the triggering of short ``bursts'' of star formation by the
environment should be quite common among dwarfs in clusters today.  
 Additionally, we have explicitly checked that mergers with other
  galaxies {\it within} the clusters are far too infrequent to be responsible
  for the measured star formation increase in cluster dwarfs. At any
  given time, satellite dwarfs have a median merger rate at least
  $\sim 1$ dex below that of the field counterparts.

 We quantify the presence of bursts of star formation by measuring
the standard deviation of the SFR at each snapshot with respect to the
time-average:

\begin{equation} 
\sigma_{\rm SFR} = \frac{\sqrt{\Sigma (\rm SFR(t)-\overline{\rm SFR})^2/N_t}}{\overline{\rm SFR}},
\end{equation}

\noindent where $N_t$ is the number of time points where we measure the
SFR  (i.e., the number of output snapshots of the simulation),
and $\overline{\rm SFR}$ is the time averaged SFR from all snapshots (indicated with a cyan curve in each
panel of Fig.~\ref{fig:sfr_panels}).  Values for $\sigma_{\rm SFR}$ are
quoted for all dwarfs in Fig.~\ref{fig:sfr_panels} and they appear to
correlate well with the visual intuition of ``burstiness'' in each
curve\footnote{We have explicitly checked the robustness of
  $\sigma_{\rm SFR}$ by first smoothing the curves with a
  Savitzky-Golay filter of width $t=0.5-1$ $\rm Gyr$. Since all
  measures seem to correlate well with the unfiltered version we
  decided to use the data without the smoothing, since the filtering
  introduces an arbitrary time-scale that is in principle unknown and
  could vary from dwarf to dwarf.}.

\begin{center} \begin{figure}
    \includegraphics[width=90mm]{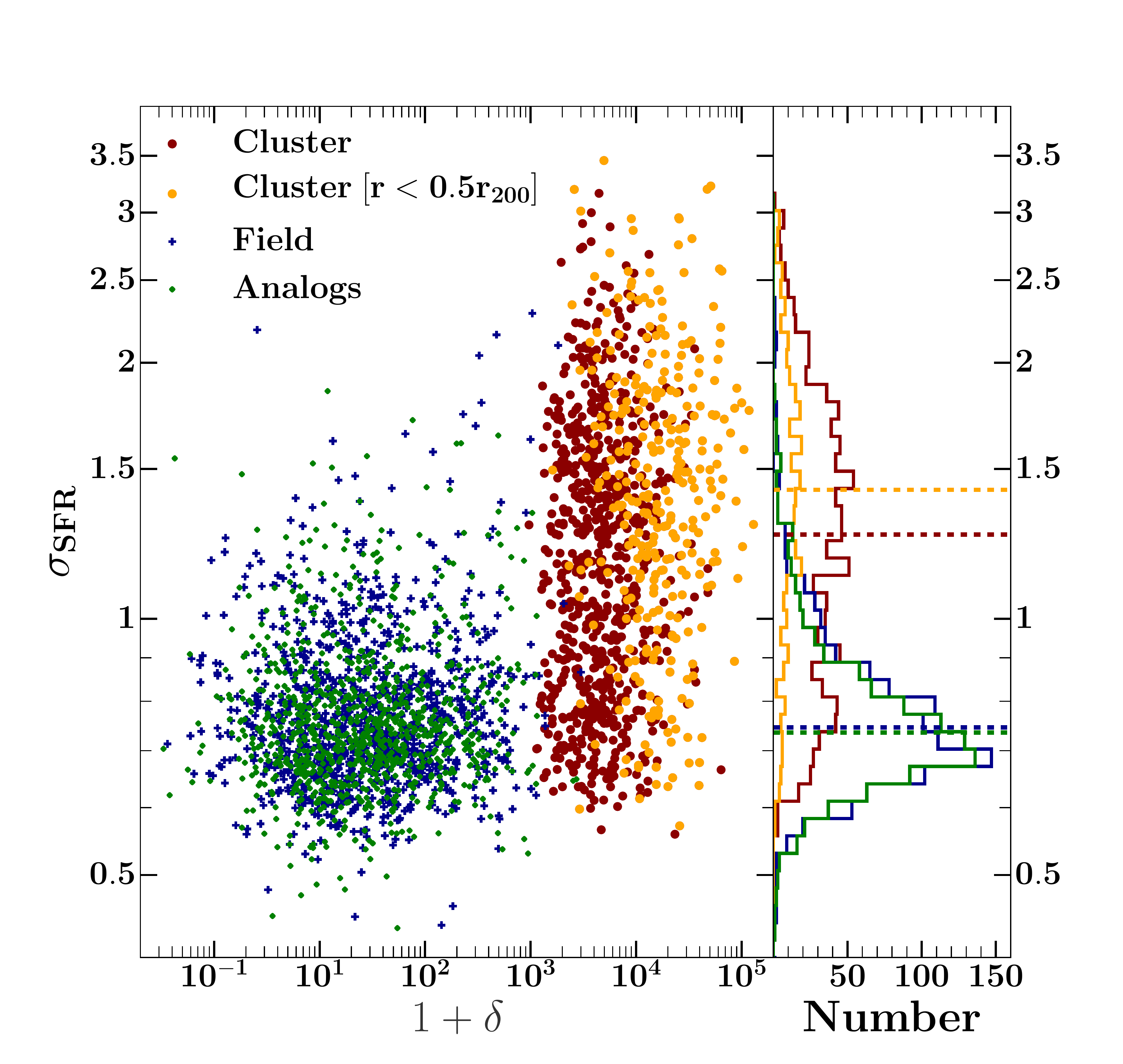}
    \caption{Measurement of ``starburstiness'' in SFRs of dwarf
      galaxies (quantified by means of $\sigma_{\rm SFR}$) as a
      function of the galaxy density of the local environment
      $\delta$.  Large $\sigma_{\rm SFR}$ are associated with the
      presence of starbursts.  Field dwarfs (blue crosses) inhabit
      less dense environments and show lower $\sigma_{\rm SFR}$ values
      compared to cluster dwarfs (red circles). The early
        ``analogs'' of cluster dwarfs (green symbols) also show low
        $\sigma_{\rm SFR}$ consistent with field objects. Histograms 
       on the right show the distribution of $\sigma_{\rm
        SFR}$ for both samples together with the median indicated by
      dotted lines. Starbursts are more common among cluster dwarfs,
      particularly those that are today in the densest inner regions
      of clusters (orange distribution corresponds to cluster dwarfs
      within $r<0.5 r_{200}$ today).}
\label{fig:rms_delta}
\end{figure}
\end{center}

Fig.~\ref{fig:rms_delta} extends the trend hinted by these 5
individual examples to all dwarfs in our samples. We show $\sigma_{\rm
  SFR}$ for cluster (red \& orange dots) and field (blue crosses)
dwarfs, plotted as a function of the local density $\delta$ (where
$\delta$ is defined as the galaxy overdensity within the radius that
encloses the 5th nearest neighbour, and only galaxies with $r$-band
magnitude brighter than $-19.5$ are used to compute the density field). Not
only do field and cluster dwarfs separate cleanly in environment (as
expected), they also show very different distributions of $\sigma_{\rm
  SFR}$. Cluster dwarfs typically have larger $\sigma_{\rm SFR}$
values, indicating the presence of significant peaks or starburst
events in their star-formation histories.

Moreover, green dots/histogram in Fig.~\ref{fig:rms_delta} show the
``analogs'' sample described in Sec.~\ref{sec:assembly} (selected to
have similar progenitor masses as cluster dwarfs at time of
infall). This sample is also characterized by low $\sigma_{\rm SFR}$
values alike the field galaxies, confirming that SFR enhancements in
cluster dwarfs are purely associated to environmental effects and is
not a bias in their progenitor mass.  Although the trend with
environment is weak above the field-cluster threshold, dwarfs in the
very inner regions of clusters show even larger $\sigma_{\rm SFR}$
compared to the rest of the satellites (the orange histogram includes
only satellite dwarfs that are today within $0.5 r_{\rm 200}$). {\it
  Our results indicate that cluster dwarfs, particularly those in the
  inner parts of clusters today, likely had a more ``bursty'' star
  formation history than dwarfs in the field.}

\begin{center} \begin{figure} 
\includegraphics[width=85mm]{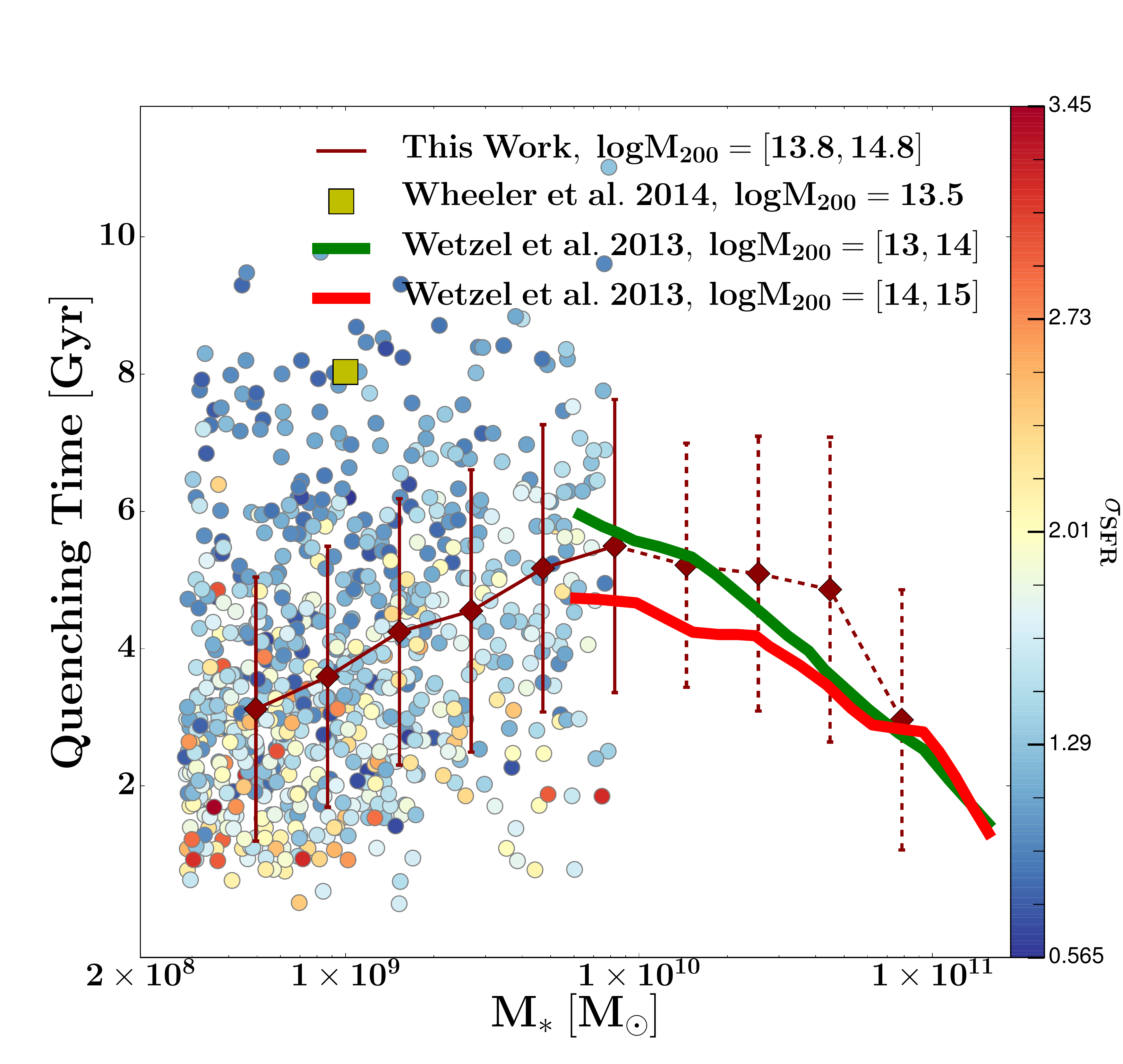}
\caption{Quenching time-scales for dwarfs in clusters as a function of
  stellar mass, where the time-scale is defined as the difference
  between infall time and the time at which the sSFR drops below
  $10^{-11}\; \rm yr^{-1}$. Cluster dwarfs are shown with circles
  color coded by their $\sigma_{\rm SFR}$. The median quenching time
  increases with stellar mass (red solid line). Interestingly,
  the dispersion around this median is large but is well-correlated with
  the {\it mode} of star formation: starbursts events (large
  $\sigma_{\rm SFR}$) are associated preferentially with the shortest
  timescales at a given stellar mass. To compare with observations,  
  we extend this analysis to satellite galaxies in
  clusters but that are more massive than our dwarfs (median and
  dispersion shown with red dashed lines). These agree well the
  time-scales derived for SDSS galaxies in Wetzel et al. 2013
  (red/green lines) for this mass range. The black square shows the
  inferred quenching time-scale for $10^9\; \rm M_\odot$ dwarfs
  orbiting around $L_*$ isolated primaries.} 
\label{fig:quench}
\end{figure}
\end{center}

Star formation in satellite galaxies is expected to be truncated
sometime after infall. Because the time-scale for this to occur sheds
valuable information on the mechanisms responsible for this quenching
(and how it may depend on host and satellite properties), the subject
has received significant attention from the community \citep[e.g.,
][]{Wang2007, Font2008,Wetzel2013, Wheeler2014, Fillingham2015}.  In
Fig.~\ref{fig:quench} we consider the time elapsed between infall and
``quenching'', defined as the first time a dwarf's SFR drops below
$\rm sSFR \approx 10^{-11} yr^{-1}$. This threshold was chosen to be
compatible with the definition in \citealt{Wetzel2013}, and we have
additionally confirmed that it applies to the scale of dwarfs by
checking that it is always $2$ dex below the average sSFR of dwarf
centrals at any given time.  For cluster dwarfs, we show the quenching
time as a function of satellite mass, where dots correspond to our
simulated dwarfs that are quiescent by $z=0$ (cluster dwarfs that are
still star-forming today do not appear in this plot; they amount to
$\sim$ a third of the sample).

Interestingly, we find a positive trend with mass: the quenching
timescale increases with the stellar mass from a median of
$\sim 3$ Gyr for our smallest objects to $\sim 5.5$ Gyr at
$M_*=10^{10} \; \rm M_\odot$ (solid dark red curve). Notice that the
scatter about this mean trend is significant, comfortably spanning the
range $\sim 1$-$8$ Gyr after infall.  Moreover, the dispersion is not random, but
appears to correlate strongly with the mode of star formation, as
quantified by $\sigma_{\rm SFR}$: dwarfs with significant
``burstiness'' (large $\sigma_{\rm SFR}$ values) preferentially show
shorter time-scales for quenching than objects with a lower
$\sigma_{\rm SFR}$ at a given stellar mass (see color-coding). Long
quenching time-scales are associated with dwarfs that passively
consume their gas while rapid quenching is often
triggered by a last
burst of star formation that consumes all available fuel for the formation of stars.

Comparing these results to observational data is not
straightforward. Samples of groups/clusters are typically not complete
down to the masses of the dwarf galaxies studied here. On the other hand,
the properties of dwarf galaxies in this mass range are well-studied
in the Local Volume, where they are easier to observe due to their
faint luminosities, or around $\sim L_*$ {\it isolated} primaries
where statistics are better for the hosts compared to more massive
groups and clusters. Although not ideal, we attempt to put our results
in the context of available observations in Fig.~\ref{fig:quench}.

We start by comparing with the quenching time-scales in clusters but
for more massive galaxies than our dwarfs (where there is
observational data). For this, we extended our
analysis to all satellite galaxies in the $12$ simulated cluster hosts
in Illustris. We select all surviving satellites at $z=0$ that are
within the virial radii of our clusters and define their quenching
time-scales in the same way as earlier. The median values obtained are
indicated with the dark red dashed curves in
Fig.~\ref{fig:quench}. For such massive satellites, we overplot
observational estimates from \citet{Wetzel2013} using SDSS data. The
agreement is good, as shown by the solid red/green curves that bracket
our host halo masses. Unfortunately, this does not guarantee that the
time-scales for lower mass dwarfs in the simulations is
reasonable; but it is
reassuring that the model is similar to the
observational data in the mass range where they overlap.

A closer look at Fig.~\ref{fig:quench} reveals a preferred mass around $M_*\sim
8 \times 10^9\; \rm M_\odot$ where this quenching timescale is maximum,
indicated by the medians in solid and dashed dark red curves. The
reason for this upturn is unclear, but is likely connected to the
physical mechanisms quenching these galaxies; we defer to a detailed
study of this in a companion paper (Mistani et al., {\it  in-prep}).
The timescales for our $M_* \sim 10^9 \; \rm M_\odot$
objects seem short compared to similar mass objects observed around
$L*$ galaxies (data taken from SDSS data in \citet{Fillingham2015} and
\citet{Wheeler2014}); however the comparison is inconclusive, as we
cannot rule out a dependence on host halo mass on this scales.
A more detailed comparison awaits until observational constraints on
the quenching of dwarfs in clusters become available.

Summarizing, our results indicate that the star formation histories of
cluster and field galaxies --with similar stellar mass today-- differ:
cluster dwarfs form their stars earlier, at higher star-formation rates and with occasional
starbursts as they orbit their host cluster.  In contrast, field dwarfs
form their stars at later times and with typically lower SFRs. This clearly
leaves an imprint in the stellar populations of these two sets of
dwarfs, explaining the much redder colors in the cluster dwarfs
compared to objects in the field.  Could
these different star formation histories also leave an imprint in the
globular clusters populations of dwarfs?  We investigate this in the
following Section.

\section{Implications for the excess of Globular Clusters in Dwarf
  Elliptical Galaxies}
\label{sec:GC}

There is not a clear picture for how globular clusters (GCs) form, but
their properties are consistent with being the evolved versions of the
young star clusters that are found in nearby
star-forming galaxies \citep{Ho1996, Mengel2002,
deGrijs2004}. For example, the GC population of dwarfs in the Local
Volume appears consistent with the initial mass function of young stellar
clusters, after accounting for stellar and tidal evolution
\citep{Kruijssen2012b}. Therefore we consider the possibility
that the processes
and physical conditions behind the formation of today's GCs are
the same than those driving the formation of young star clusters
at $z=0$.

%
%

Analytical arguments suggest that  gas-rich, high-pressure environments
offer the most propitious conditions for the formation of bound
star clusters \citep{Harris1994, Elmegreen1997,
Shapiro2010, Kruijssen2012}; a claim that has also found support  from
numerical  simulations \citep{Kravtsov2005, Bournaud2008,
Renaud2008}.  Observationally, there is strong evidence in favour
of a link between star cluster formation and local properties of
the interstellar medium. In particular, the fraction of stars born in clusters (that
survive the embedded phase) compared to the total amount of stars
formed, termed $\Gamma$, appears to correlate with the density of star
formation as \citep{Goddard2010}: 

\begin{equation}
\Gamma = 0.29 \bigg(\frac{\Sigma_{\rm SFR}}{M_\odot yr^{-1}kpc^{-2}}\bigg)^{0.24},
\label{eq:gamma}
\end{equation}

\noindent with $\Sigma_{\rm SFR}$ defined as the averaged projected
density of star formation in a galaxy. Such a scaling was later
confirmed by other independent studies \citep{Adamo2010, Silva2011,Adamo2011}.
 If indeed star clusters form at the high density tail
of the interstellar medium \citep{Kruijssen2012}, then it is expected that
GC formation was more efficient at higher redshifts (when galaxies
were typically more gas-rich and turbulent) and during starburst
events; interestingly this corresponds to two of the main differences in the star
formation histories of our cluster dwarfs compared to 
our field galaxies. 

To address this issue in some detail and to roughly predict the GC
efficiencies in cluster and field dwarfs, we implement a simple model
for the formation of GCs as a post-processing algorithm in our simulations. The
model involves three steps: $i)$ compute the amount of stellar mass
born in star clusters, $ii)$ sample an initial mass function of
clusters and $iii)$ use a particle tagging scheme to address the tidal
removal of GCs from dwarfs orbiting inside galaxy clusters. 

For the first step, we use the empirical relation from
\citet{Goddard2010}, where for each dwarf at a given time we compute
the fraction of stars formed that are born in bound clusters using
Eq.~\ref{eq:gamma}. We define $\Sigma_{\rm SFR}$ as the sum of the
star formation of all gas cells in the galaxy and we divide by the
area of a circle with radius equal to that enclosing $95\%$ of the
star-forming gas (a good proxy for the $H_\alpha$ radius). We then
repeat this calculation throughout the time evolution of each dwarf,
averaging in time-steps of $1$ Gyr (we have explicitly checked that
the results do not depend on the time window used). Adding up the mass
in stars that is born in clusters for each of the time-steps gives us
a prediction for the total mass expected in star clusters for each
dwarf.

\begin{center} \begin{figure}
\includegraphics[width=85mm]{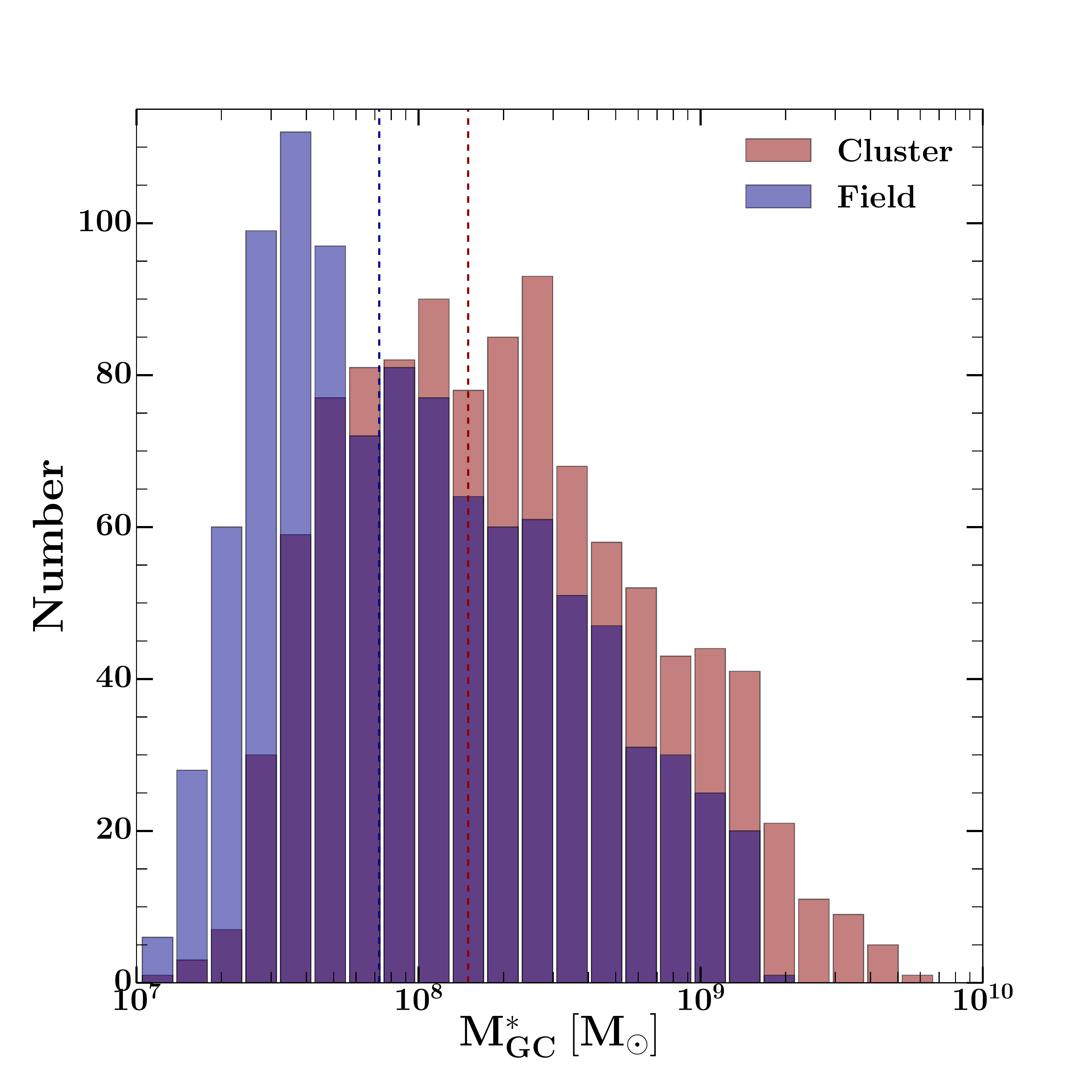}
\caption{Mass in GCs predicted for the field (blue) and cluster
  (red) dwarf samples according to our post-processing model of GC
  formation (see Sec.~\ref{sec:GC}). Although the distributions
  overlap, dwarfs in clusters have on average $\sim 2$ times more mass
  in GCs than the field dwarfs; an effect dominated by the earlier and
  more bursty star-formation in cluster dwarfs.}
\label{fig:GC_mass}
\end{figure}
\end{center}
%

%
%
\begin{center} \begin{figure}
\includegraphics[width=85mm]{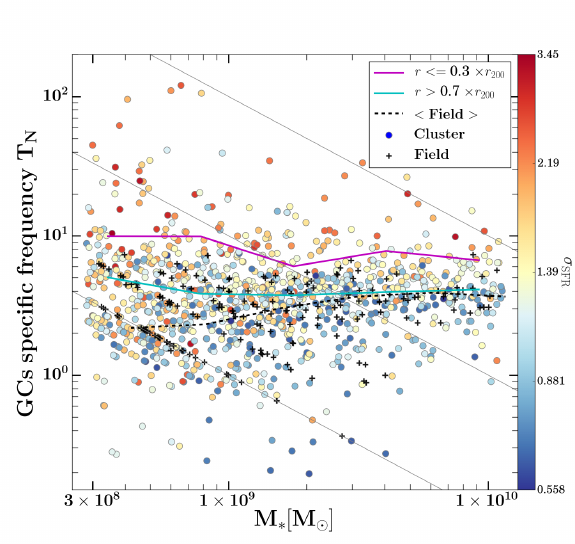}
\caption{Final specific frequency of GCs for the sample of dwarfs in
  clusters (colored circles) and in the field (black crosses). $T_N$
  is computed as the number of GCs within an aperture of $8$ kpc
  divided by the stellar mass of each object. $T_N$ also accounts for
  the effects of environmental tidal stripping. In agreement with
  observations, we find that cluster dwarfs have a larger frequency of
  GCs than field dwarfs of the same $M_*$, a trend that is enhanced
  for dwarfs with starbursty star formation histories (higher
  $\sigma_{\rm SFR}$, see color coding) and/or dwarfs in the inner
  regions of clusters. Magenta and cyan lines show the medians for
  cluster dwarfs with $r<0.3 r_{200}$ and $r>0.7 r_{200}$,
  respectively; field dwarfs shown with dotted black line. (For
  clarity, we only show 1 in 5 field dwarfs selected randomly from the
  sample.)  Inclined grey lines show lines of constant number of GCs:
  $100$, $10$ and $1$ from top to bottom.  Differences are {\it
    relative} between the field and cluster samples, as the vertical
  normalization is chosen to coincide roughly with observed $T_N$
  range (see text for more detail).  }
\label{fig:T_N}
\end{figure}
\end{center}
%

Fig.~\ref{fig:GC_mass} shows the distribution of the
computed mass in GCs for the population of cluster (red) and field
(blue) dwarfs. Although the distributions overlap, we find a
significant difference between the two, with cluster dwarfs having on
average a factor $\sim 2$ times more stellar mass in bound clusters
than field dwarfs (vertical lines show the median in each sample). The
difference is largely due to the earlier and higher SFRs in the
former, as discussed in Sec.~\ref{sec:sfr}. This overall trend is in
good agreement with results from the semi-analytical models presented in
\citet{Peng2008}, but the star-formation histories of field and
cluster dwarfs are much more alike in their model (based on the
Millennium simulations plus \citet{Guo2011} semi-analytical catalog)
than in our sample.

In order to compare more closely with observational 
measurements of the
GC efficiency in dwarfs, we need to translate our estimates of mass in
GCs to GC numbers. Following \blue{Jordan et al. (2007b)}, 
we assume a Schechter-like cluster mass
function and randomly sample from such a distribution until we
account for the total GC mass for each dwarf. In particular, we adopt: 

\begin{equation}
 \frac{dN}{dM} \propto M^{-2}exp\big(-M/M_c\big),
\end{equation}

\noindent where $M_c = 2\times10^5 M_\odot$ is the truncation
mass. For our study, we focus on the high mass end, where GCs are
more resilient to destruction by internal and environmental
mechanisms.  We sample globular clusters in a mass range of
[$1\times10^4 M_\odot$, $1\times10^8 M_\odot$], where the upper bound
is set to ensure that GCs have less mass than the galaxy itself.
This is in general not important, as the rapid fall-off of
the Schechter function ensures that high-mass clusters are highly
uncommon. Notice that the low mass end we choose for the sampling works
as a ``normalization'' factor to our total number of GCs (if we choose
a smaller mass, we get a higher number of GCs and the same mass can now
be distributed among many smaller-mass GCs than before). Thus, 
in what follows, our results
are {\it differential} between both the dwarf samples.  Our
main aim is to compare the results for
the cluster and field samples with respect to each other
under the same assumptions.

For dwarfs in the field, this is enough to compute their GC specific
frequency $T_N$, defined as the number of GCs divided by the stellar
mass of the dwarf, $T_N=N_{gc}/M_*/10^9\; \rm M_\odot$
\citep{Zepf1993}.  However, for dwarfs in clusters, the tidal forces
exerted by the host cluster potential can potentially strip some of
the GCs originally formed in-situ. We therefore implement a
particle-tagging technique, where we ``paint'' GCs on some dark-matter
particles and use those particles as tracers of the time evolution for
the GC population. This technique is not dynamically self-consistent
(the mass of the dark-matter particle does not coincide with the GC
that was ``painted'' onto it) but the expected error is small. The
technique has been successfully used before in studies of GCs 
\citep{Griffen2010, Ramos2015}, stellar halos \citep{Cooper2010, Cooper2013},
brightest cluster galaxies \citep{Napolitano2003, Laporte2012, Cooper2015a}, and structure of dwarf
spheroidal galaxies \citep{Penarrubia2008}, among others.

For our tagging technique, we randomly select particles following a
steep density profile, $\rho_{\rm GC} \propto r^{-3.5}$, which is
consistent with the distribution of GCs around the Milky Way according
to the GC catalog in \citet{Harris1996}. We use a maximum distance
equal to $10$ kpc, but the steep slope ensures that the half-number
radius is only $1$-$2$ kpc, in agreement with observations of dwarfs
\citep{Georgiev2009b}.  Our results do not depend on this maximum
radius assumed. For each cluster dwarf, the tagging in time is done
only once, at $t_{\rm max}$.  Although some of the star clusters will
continue to form after infall into the galaxy cluster, we assume that
the total number of GCs is already in place at the time of infall,
which simplifies the procedure compared to tagging particles at
individual time-steps. This approach is conservative, as tagging
particles at an earlier time can only increase the chances of
stripping later on.

After the particles have been selected at $t_{\rm max}$ for each
dwarf, we use their IDs to check for their positions at $z=0$ and
consider the GCs as bound/stripped according to their distance to the
dwarf where it was born. In order to approximate the aperture used in
studies of the Virgo cluster (\blue{Jordan et al.  2007b, Peng et
  al. 2008}), we use $r_{\rm cut}=8$ kpc, but have also checked that
apertures in the range $5$-$10$ kpc yield similar results.  For field
dwarfs (not exposed to stripping) the tagging is not necessary, and we
simply do the spatial sampling once at $t_{\rm max}$ and check for the
number of GCs within the aperture.  In order to average-out the
results of a particular random realization, we repeated the tagging
exercise $10$ times for each dwarf and assigned the final $T_N$ based
on the average of these $10$ realizations.

Fig.~\ref{fig:T_N} shows the final specific frequency of GCs, $T_N$,
predicted by our model using all massive GCs with mass above $M =4
\times 10^5 \; \rm M_\odot$ (this mass cut-off is not important, but
was chosen to agree roughly with the average normalization of observed GC
frequencies in dwarfs). We show the predicted $T_N$ as a function of
stellar mass for the sample of cluster (solid colored dots) and field
(black crosses) dwarfs. We find a clear excess of GCs for cluster
dwarfs, even after accounting for tidal stripping within the host
clusters, with the median $T_N$ of cluster dwarfs about $\sim 1.65$
times larger than dwarfs in the field. Most importantly, we find a clear
correlation between this GC excess and the ``starburstiness'' of the
star-formation histories in the dwarfs, as shown by the color coding
of the dots: cluster dwarfs with clear peaks in their star formation
histories (larger $\sigma_{\rm SFR}$) preferentially dominate the tail
of high GC frequencies, whereas dwarfs with a more quiet star
formation history tend to show $T_N$ closer to those of field dwarfs.
{\it We therefore see that the combination of early star formation
  plus the starburst events associated with infall and pericentric
  passages for cluster dwarfs can account for the more numerous GC
  populations observed for dEs.}

%
\begin{center} \begin{figure*} 
\includegraphics[width=0.85\linewidth]{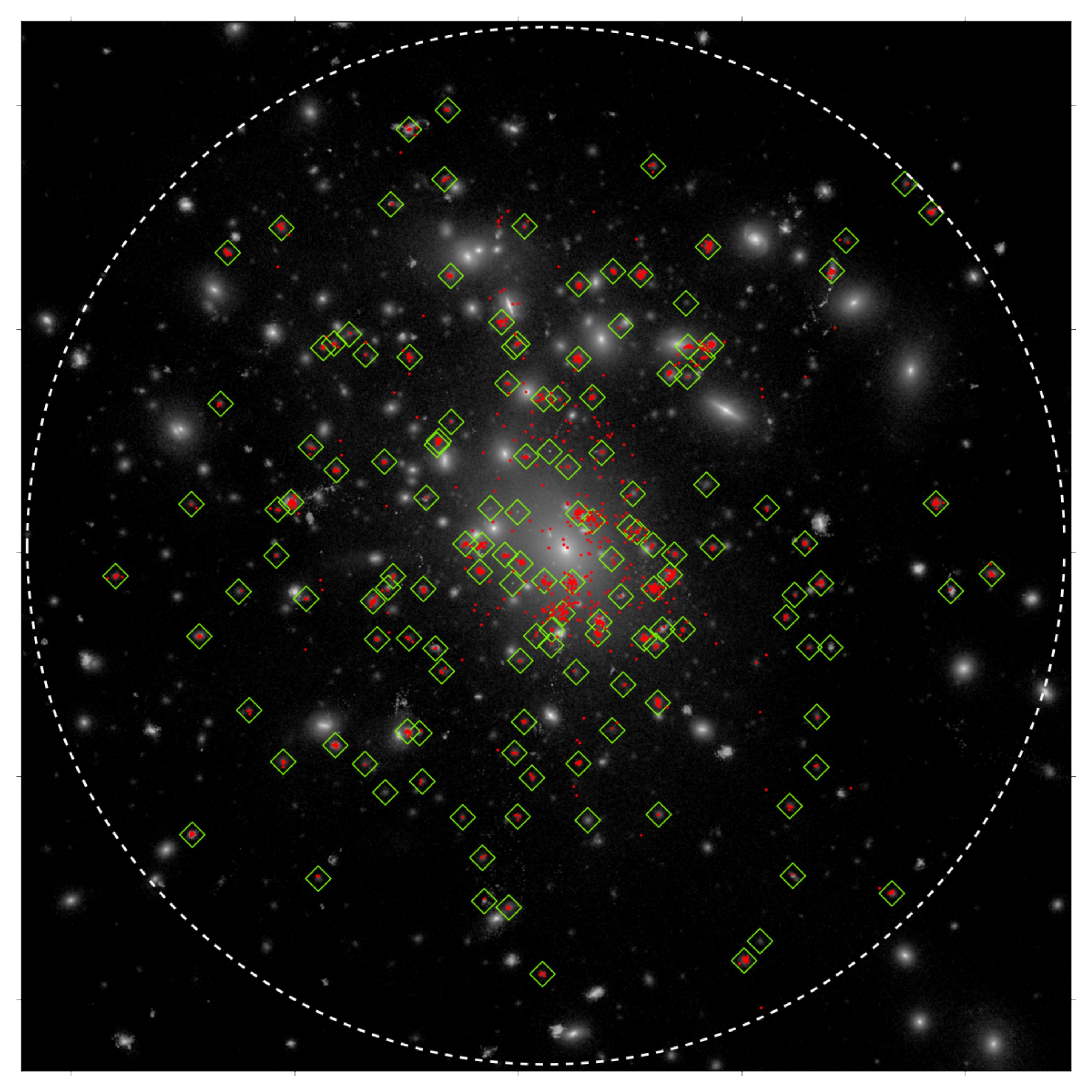}
\caption{Final distribution of tagged GCs (red dots) overlaid on the
  projected stellar-mass map of our most massive cluster. Dwarf
  galaxies in this cluster (the only contributors of GCs in this
  figure) are highlighted with green squares. Although most of the GCs
  remain bound to the dwarfs where they formed, several of them have
  been tidally stripped from their native dwarf galaxy, particularly in
  the inner regions of the cluster where tidal forces are
  stronger. The stellar content and phase-space properties (positions
  and velocities) of such stripped GCs could still retain information
  of their dwarf progenitors, such as orbits, star-formation histories
  and also help determine the tidal radii of cluster dwarfs.  }
\label{fig:map}
\end{figure*}
\end{center}
%

Observationally, much of the excess in GC specific frequency comes
from nucleated dEs whereas non-nucleated dEs have GC
frequencies more similar to those of dIrrs \citep{Miller1998,MillerLotz2007}. 
Although our numerical resolution is not sufficient to
address the morphology of the simulated dwarfs, the correlation
between high $T_N$ and the presence of starbursts suggests that
nucleated dEs are simply the tail of the distribution of dwarf
galaxies where environment caused the strongest bursts of
star-formation associated with infall and pericentric passages, in
agreement with the proposal of \citet{MillerLotz2007}. Moreover, the excess
in $T_N$ found in Fig.~\ref{fig:T_N} is dominated by dwarfs located in
the inner regions of the clusters. The magenta line shows the median
specific frequency of GCs for dwarfs with $r<0.3 r_{200}$ which is a
factor $\sim 1.5$ larger than that of cluster dwarfs in the outskirts $r>0.7
r_{200}$ (cyan curve) and dwarfs in the field (dashed black curve). This
radial dependence is in agreement with the observed trend
for dEs in the Virgo cluster \citep{Peng2008}. 

The effects of tidal stripping can be significant, but is not enough
to erase the initial difference (Fig.~\ref{fig:GC_mass}) expected from
our calculation of the mass in GCs.
Fig.~\ref{fig:map} shows a more comprehensive view of the final
distribution of GCs (red dots) with respect to the dwarfs (green
squares) within a galaxy cluster (in this case, the most massive of
our host clusters). The image shows a luminosity-weighted stellar mass
map and indicates that although the majority of GCs are still bound to
the dwarfs in which they were born, there are several that have been
stripped by the cluster potential and may now belong to the background
population of GCs of the massive central galaxy in the cluster. Dwarf
elliptical galaxies seem then able to contribute a fair number of
``accreted'' GCs to the central potential of the host cluster.
 Moreover, tidally disrupted and unresolved GCs will contribute to
the diffuse intra-cluster light instead, a study of which we defer to
future work.

The results of this section are encouraging, but we emphasize the
limitations of our model. One strong assumption we have made is that
the star cluster efficiency $\Gamma$ scales, {\it at all redshifts} and
for low mass galaxies as the one considered here, with the density of
star formation as indicated by Eq.~\ref{eq:gamma}. The equation is a
power-law fit to observations in nearby star-forming galaxies, and it
is not guaranteed that the same relation holds at higher
redshifts.  In order to explore alternative parametrizations of
$\Gamma$, we considered the theoretical models presented in
\citet{Kruijssen2012}, which are based on physical properties of the
gas and therefore can be applied at any given time. In this model,
$\Gamma$ scales with the surface density of the gas $\Sigma_g$, which
we can measure from our simulations.

Although the \citeauthor{Kruijssen2012} model is more detailed than our numerical
resolution allows (it requires measurements for the stability
parameter $Q$ for galaxy disks, among other factors), using standard values
of $Q=1.5$, we were able to reproduce the factor $\sim
2$ relative excess in GC mass for dwarfs in clusters compared to the
field\footnote{We thank the author for making the code to compute GC
  efficiency publicly available at
  http://www.mpa-garching.mpg.de/cfe.}. Additionally, we have checked that
our definition of size used to compute $\Sigma_{\rm SFR}$ in
Eq.~\ref{eq:gamma} does not change the results we presented. To do
this we vary the definition of size using: the radius enclosing $95\%$
of the star forming gas (the proposed model), half-mass radius of the
stars and half-mass radius of the gas, and in all cases recover
the {\it relative} difference between cluster and field dwarfs
presented in Fig.~\ref{fig:T_N}.

A second important assumption in our model is that the destruction of
massive GCs is similar for dwarfs in clusters and in the
field. However, for massive GCs, dynamical friction (DF) forces can
accelerate their orbital decay, eventually causing them to merge with
the host galaxy. Analytical estimates of time needed for GCs to sink
to the center of their hosts due to DF can vary from $1$-$10$ Gyr
depending on GC mass for a host dwarf galaxy with $M_* \sim 10^9\; \rm
M_\odot$ \citep[see fig. 19 in ][]{Turner2012}. This indicates that the
magnitude of the GC excess in cluster dwarfs can be overestimated in
our model. Nonetheless, the net effect of DF is far from clear. For
example, the DF timescale depends strongly on the (unknown) initial
distribution of the GCs and also on the dark matter density profile of
the dwarfs. Halos with lower central densities or constant dark matter
cores show much longer DF timescales \citep{Hernandez1998,
  Cole2012}. Because tidal stripping of the dark matter halo in
satellite dwarfs will tend to lower the central densities
\citep[e.g. ][]{Hayashi2003}, the sinking of GCs in cluster dwarfs
could result, overall, less efficient than in the field. Therefore,
although some of the most massive GCs in cluster dwarfs may merge with
their hosts, due to the steep mass function of GCs favoring {\it in
  number} lower mass GCs, the GCs frequency in cluster dwarfs is
likely to remain significantly enhanced with respect to the field.

Finally, we also tested the effects of assuming a shallower GC radial
distribution (and thus more prone to tidal stripping for cluster
dwarfs). Following observational results by \citet{Beasley2009}, we
assumed $\rho \propto r^{-2}$ (instead of the steeper $-3.5$ slope
found in the Milky Way) and sample the position of GCs within a
maximum radius of $5$ kpc. This cut-off was chosen such that the
radius containing half of the GCs is $\sim 1$-$2$ kpc as observed for
dwarfs. We repeated the particle tagging step
and check for the number of GCs that remained bound to their host
dwarfs, as before. We find that although some cluster dwarfs are
stripped more easily of their GCs, the majority of the sample manages
to retain most of their GCs, suggesting that our result in
Fig.~\ref{fig:T_N} is robust to changes in the radial distribution of
GCs that are within observational constraints \citep[see also ][]{Smith2013}.

We conclude that the model is not overly sensitive to the arbitrary
choices we have adopted and that it represents an attractive tool for
studying the formation of GCs in cosmological simulations of galaxies
where the resolution is not adequate to describe the scale of
individual star clusters. The application of this model to our sample
of dwarf galaxies indicates that dwarf ellipticals in clusters would
naturally have an enhanced frequency of GCs compared to field dwarfs,
an effect that is most prominent for dwarfs in the inner parts of
clusters; and that such a difference is not only a reflection of an
early assembly but is also due to the starburst events associated with
infall into the host cluster.


\section{Summary \& Conclusion}
\label{sec:concl}

We use the Illustris simulation to study the formation of dwarf
galaxies in clusters.  We select $12$ host galaxy clusters in the
virial mass range $M_{\rm 200}=0.72$-$2.32 \times 10^{14}\; \rm
M_\odot$, with the high mass end akin to systems like the Virgo and
Fornax clusters.  Dwarf galaxies are selected to be within the virial
radii of these clusters at $z=0$ and have stellar masses in the range
$M_*=3\times 10^8$-$1\times 10^{10}\; \rm M_\odot$. We find 1071
candidate dwarfs that satisfy our selection criteria. We construct an
equal-number control sample of field dwarfs selected in the same
stellar mass range at $z=0$ but that are central objects of their FoF
group. Our main findings can be summarized as follows:

\begin{itemize}

\item At $z=0$ the population of dwarfs in clusters has redder colors
  and lower star formation rates than field objects, in good agreement
  with the properties of observed dEs. The parameters in the
  simulation were not tuned to reproduce this, so this is a successful 
  prediction of our model that motivates further studies into the
  evolution that shaped these two different populations.

\item Dwarfs in clusters are mostly genuine low mass
  objects and not the descendants of more massive
  galaxies
  that have been substantially stripped. 
  Only $14\%$ of our dwarfs have lost more than $50\%$ of
  their stellar mass due to tidal effects. 

\item Although by construction both samples have the same stellar mass
  at $z=0$, the assembly of cluster and field dwarfs is
  remarkably different: cluster dwarfs form their stars at earlier
  times and at higher star formation rates compared to field
  dwarfs. Because of this difference in the build-up times, the
  progenitors of dwarfs that are in clusters today were, at the average infall
  time of the sample $z \sim 1$, about $3.5$ times more massive than
  the progenitors of today's field dwarfs at the same time. 

\item The star formation histories of dwarfs in clusters often show
  significant starburst events which are not present in the population
  of field dwarfs. These starbursts are associated with infall and/or
  first pericentric passages, likely due to gas compression in
  the dwarfs under the combined effects of tidal and ram-pressure
  forces.

\item The cessation of star formation in cluster dwarfs occurs within
  timescales (median) $\sim 3$-$5.5$ Gyr after they become satellite
  galaxies.  We find a positive trend with mass, in which low mass dwarfs go
  quiescent most rapidly. Nevertheless, there is a large
  scatter around these median values, with differences $\ge 7$ Gyr at
  the same $M_*$. This highlights the varied paths by which
  star formation is halted: dwarfs with significant starburst events
  quench rapidly, whereas others in the sample simply consume their
  gas fuel and quench over much longer timescales.

\item The different star formation histories of cluster and field
  dwarfs can have a significant impact on their populations of globular
  clusters. To study this, we implement a post-processing model of GC
  formation and apply it to our sample of cluster and field dwarfs. In
  a scenario where: $a)$ GCs are the evolved population of young
  star clusters and $b)$ the fraction of stars born in clusters
  compared to field stars scales with either the star formation
  density or gas density, our results indicate that dwarfs living in
  galaxy clusters today should have an enhanced number of globular
  clusters per unit stellar mass compared to galaxies in the field
  with similar stellar mass, in agreement with
  observations. After accounting for stripping, we find that the
  difference in the mean specific frequency of GCs, $T_N$, is a factor $\sim 1.6$, with extreme cases
  having $T_N$ as much as $10$ times larger than the field dwarf
  population.
\end{itemize}

Our results lend support to the notion that dEs in clusters share a
similar origin with isolated dIrr galaxies in the field, but their
properties have been fundamentally altered due to the cluster
environment.  This idea, although perhaps the most popular scenario for
the formation of dEs, appeared to be in tension with a larger specific
frequency of GCs observed for cluster dwarfs.  Here we find that such
a trend could arise from the different star formation histories of the
samples.

We emphasize, however, that the {\it progenitors} of cluster and field
dwarfs of similar mass at $z=0$ can differ substantially in their
masses at earlier times.  Observational studies that are seeking to
find a population of isolated dwarfs with the same {\it progenitors}
as dwarfs in clusters today, may need to focus on present-day
  gas-rich galaxies that are $\sim 3$ times more massive than the
  target dE population.  This difference corresponds to the
freeze-out of star formation that occurs for cluster dwarfs after
infall, while the analog field population can continue to form stars
at a normal rate outgrowing their cluster peers from $z \sim 1$ until
now.

The results presented here have a number of predictions
that can be tested observationally. For example, in our simulations a
significant number of cluster dwarfs show intense starburst episodes,
the majority of which are then followed by a rapid shut-off of star
formation. Such events will leave a clear imprint in the stellar
populations of cluster dwarfs, likely with $\alpha$-element
enhancements compared to dwarfs in the field where star formation
proceeded more slowly.  Such a trend could be seen in observations when
detailed metallicity data becomes available for dEs in nearby
clusters. 

Another interesting prediction pertains to the kinematics of GCs
around dEs.  Our modeling of GCs suggests that some of the native GCs
in dwarfs may have been stripped (or are in the process of being
stripped) by the host potential.  These stripped GCs will retain
information about their origin imprinted in their phase-space
properties (positions and velocities) that could be used to design
observational strategies to identify the full population of native GCs
born in dE galaxies.

Finally, current and upcoming observations provide valuable insight
into the structural properties of cluster dwarfs. Results so far have
revealed a fascinating variety, with dEs showing little or no
rotation, well defined but faded disks, and even counter-rotating
components (\blue{Toloba et al 2014a,b}). Moreover, dEs in the Local
Group are also starting to show interesting differences in their star
formation histories that could be linked to their different orbits
\citep{Geha2015}. In a follow-up study, we will focus on the analysis of
zoom-in simulations of clusters that have the resolution needed to
address the morphological transformation of dwarfs in clusters. This
will provide a comprehensive theoretical view of the evolutionary link
between dEs and their suspected dIrr progenitors and the many paths
that have helped to shape this dichotomy in the dwarf galaxy
population.

%
%
\section*{Acknowledgements}
The authors are grateful to Mario Abadi, James Bullock, Mike Cooper,
Raja Guhathakurtha, Julio Navarro, Diederik Kruijssen, Eric Peng,
Elisa Toloba and Carlos Vera-Ciro for insightful and stimulating
discussions.  We would like to thank Shy Genel, Debora Sijacki and
Volker Springel for early access to the simulations and comments on
the draft. We would like to thank the referee, Hugo Martel, for 
helpful comments that improved the manuscript.
Simulations were run on the Harvard Odyssey and CfA/ITC
clusters, the Ranger and Stampede supercomputers at the Texas Advanced
Computing Center as part of XSEDE, the Kraken supercomputer at Oak
Ridge National Laboratory as part of XSEDE, the CURIE supercomputer at
CEA/France as part of PRACE project RA0844, and the SuperMUC computer
at the Leibniz Computing Centre, Germany, as part of project pr85je.
L.H. acknowledges support from NASA grant NNX12AC67G and NSF grant
AST-1312095. AP acknowledges support from the HST grant HST-AR-13897.


\bibliography{master}
\end{document}